\documentclass[english]{sig-alternate}

\pdfoutput=1
 
\usepackage[latin9]{inputenc}
\usepackage{color}
\usepackage{float}
\usepackage{amsmath}
\usepackage{amssymb}
\usepackage{graphicx}
\usepackage{booktabs}
\usepackage{hyperref}

\floatstyle{ruled}
\newfloat{algorithm}{tbp}{loa}
\providecommand{\algorithmname}{Algorithm}
\floatname{algorithm}{\protect\algorithmname}

\usepackage{times}
\usepackage{color}
\definecolor{gray}{gray}{0.96} 

\usepackage[caption=false]{subfig}

\usepackage{listings}

\makeatletter
\def\@copyrightspace{\relax}
\makeatother
 
\begin{document}

\numberofauthors{2} 
\author{ 
	\alignauthor 
	Patrick Schäfer\\        
	\affaddr{Humboldt University of Berlin}\\        
	\affaddr{Berlin, Germany}\\        
	\email{patrick.schaefer@hu-berlin.de} 
	\alignauthor 
	Ulf Leser\\        
	\affaddr{Humboldt University of Berlin}\\        
	\affaddr{Berlin, Germany}\\        
	\email{leser@informatik.hu-berlin.de}
}

%
%

\title{Fast and Accurate Time Series Classification with WEASEL}
\maketitle
\begin{abstract}
Time series (TS) occur in many scientific and commercial applications, ranging from earth surveillance to industry automation to the smart grids. An important type of TS analysis is classification, which can, for instance, improve energy load forecasting in smart grids by detecting the types of electronic devices based on their energy consumption profiles recorded by automatic sensors. Such sensor-driven applications are very often characterized by (a) very long TS and (b) very large TS datasets needing classification. However, current methods to time series classification (TSC) cannot cope with such data volumes at acceptable accuracy; they are either scalable but offer only inferior classification quality, or they achieve state-of-the-art classification quality but cannot scale to large data volumes.

In this paper, we present WEASEL (Word ExtrAction for time SEries cLassification), a novel TSC method which is both scalable and accurate. Like other state-of-the-art TSC methods, WEASEL transforms time series into feature vectors, using a sliding-window approach, which are then analyzed through a machine learning classifier. The novelty of WEASEL lies in its specific method for deriving features, resulting in a much smaller yet much more discriminative feature set. 
On the popular UCR benchmark of 85 TS datasets, WEASEL is more accurate than the best current non-ensemble algorithms at orders-of-magnitude lower classification and training times, and it is almost as accurate as ensemble classifiers, whose computational complexity makes them inapplicable even for mid-size datasets. The outstanding robustness of WEASEL is also confirmed by experiments on two real smart grid datasets, where it out-of-the-box achieves almost the same accuracy as highly tuned, domain-specific methods.
\end{abstract}

\sloppy


\maketitle

\section{Introduction}
A (one-dimensional) time series (TS) is a collection of values sequentially ordered in time. TS emerge in many scientific and commercial applications, like weather observations, wind energy forecasting, industry automation, mobility tracking, etc. One driving force behind their rising importance is the sharply increasing use of sensors for automatic and high resolution monitoring in domains like smart homes~\cite{jerzak2014debs}, starlight observations~\cite{protopapas2006finding}, machine surveillance~\cite{mutschler2013debs}, or smart grids~\cite{WindPower, hobbs1999analysis}. 

Research in TS is diverse and covers topics like storage, compression, clustering, etc.; see~\cite{esling2012time} for a survey. In this work, we study the problem of time series classification (TSC): Given a concrete TS, the task is to determine to which of a set of predefined classes this TS belongs to, the classes typically being characterized by a set of training examples. Research in TSC has a long tradition~\cite{bagnall2016great,esling2012time}, yet progress was focused on improving classification accuracy and mostly neglected scalability, i.e., the applicability in areas with very many and/or very long TS. However, many of today's sensor-driven applications have to deal with exactly these data, which makes methods futile that do not scale, irrespective of their quality on small datasets. Instead, TSC methods are required that are both very fast and very accurate.

\begin{figure}
\includegraphics[width=1\columnwidth]{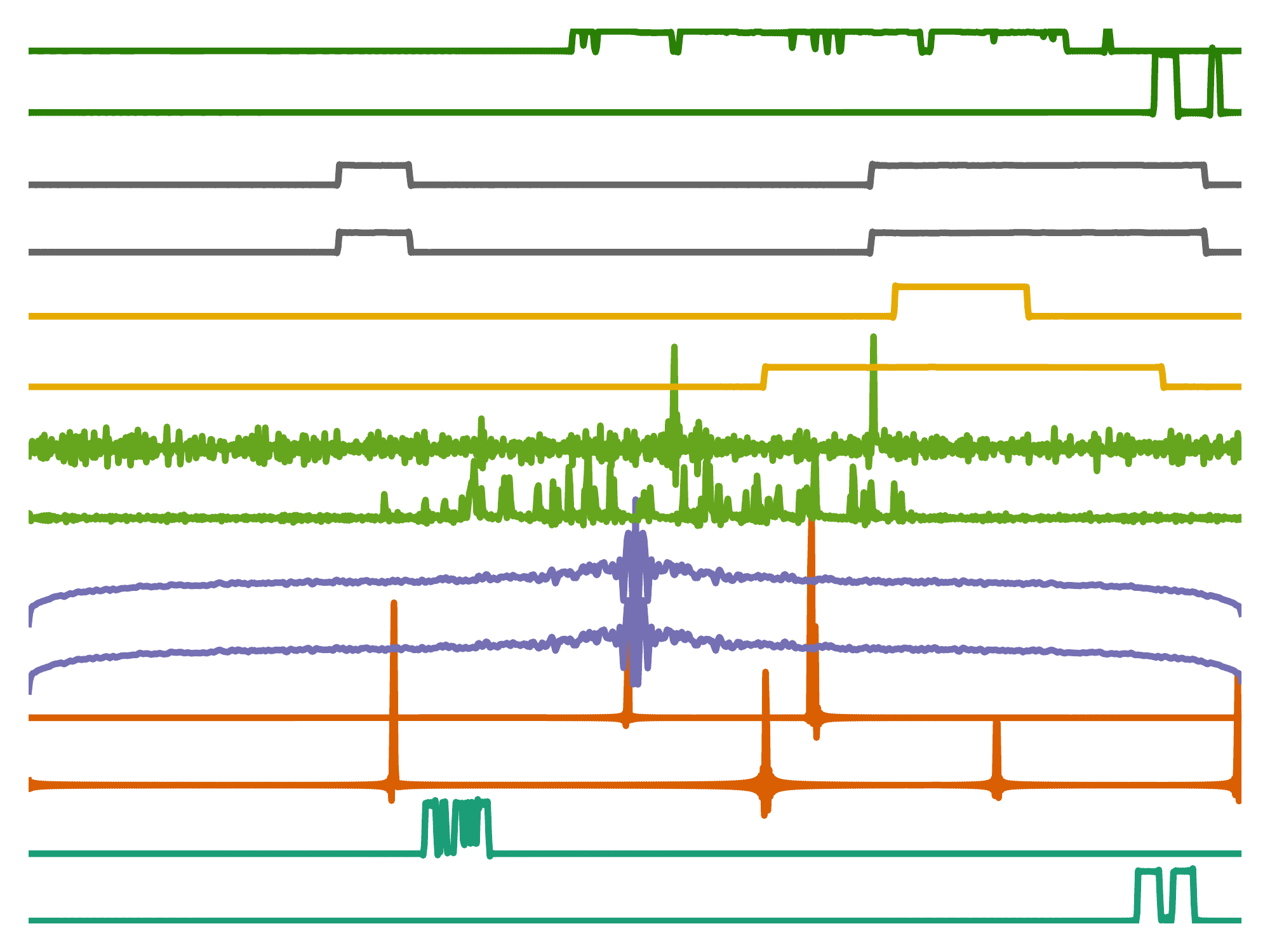}

\caption{Daily power consumption of seven appliances with two samples per class. Bottom to top: dishwasher, microwave oven, digital receiver, coffee-maker, amplifier, lamp, monitor.\label{fig:DailyPowerConsumption}}
\end{figure}

As a concrete example, consider the problem of classifying energy consumption profiles of home devices (a dish washer, a washing machine, a toaster etc.). In smart grids, every device produces a unique profile as it consumes energy over time; profiles are unequal between different types of devices, but rather similar for devices of the same type (see Figure~\ref{fig:DailyPowerConsumption}). The resulting TSC problem is as follows: Given an energy consumption profile (which is a TS), determine the device type based on a set of exemplary profiles per type. For an energy company such information helps to improve the prediction of future energy consumption~\cite{gisler2013appliance,gao2014plaid}. 
For approaching these kinds of problems, algorithms that are very fast and very accurate are required. Regarding \textbf{scalability}, consider millions of customers each having dozens of devices, each recording one measurement per second. To improve forecasting, several millions of classifications of time series have to be performed every hour, each considering thousands of measurements. Even when optimizations like TS sampling or adaptive re-classification intervals are used, the number of classifications remains overwhelming and can only be approached with very fast TSC methods.
Regarding \textbf{accuracy}, it should be considered that any improvement in prediction accuracy may directly transform into substantial monetary savings. For instance,~\cite{WindPower, hobbs1999analysis} report that a small improvement in accuracy (below 10\%) can save tens of millions of dollars per year and company. However, achieving high accuracy classification of home device energy profiles is non trivial due to different usage rhythms (e.g., where in a dishwasher cycle has the TS been recorded?), differences in the profiles between concrete devices of the same type, and noise within the measurements, for instance because of the usage of cheap sensors.

\begin{figure}
\includegraphics[width=1\columnwidth]{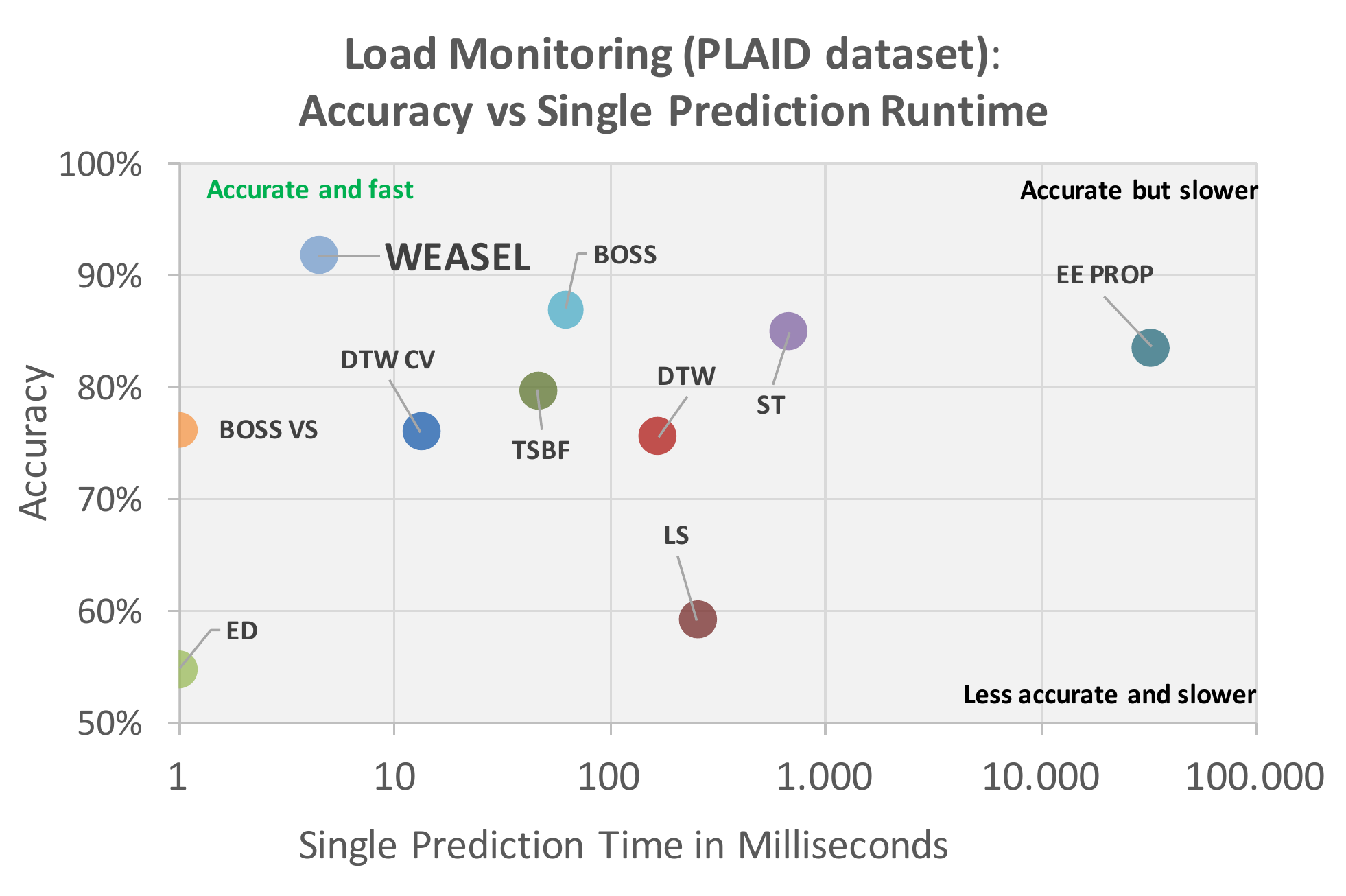}

\caption{Classification accuracy and single prediction runtime (log scale) for different TSC methods on the energy consumption dataset PLAID. Runtimes include all preprocessing steps (feature extraction, etc.). Methods are explained in detail in Section 2, the system used for measurements is described in Section 5.\label{fig:performance_on_plaid}}
\end{figure}

Current TSC methods are not able to deal with such data at sufficient accuracy and speed. Several high accuracy classifiers, such as Shapelet Transform (ST)~\cite{bostrom2015binary}, have bi-quadratic complexity (power of $4$) in the length of the TS; even methods with quadratic classification complexity are infeasible. The current most accurate method (COTE~\cite{bagnalltime}) even is an ensemble of dozens of core classifiers many of which have a quadratic, cubic or bi-quadratic complexity. On the other hand, fast TSC methods, such as BOSS VS~\cite{schafer2015scalable2} or Fast Shapelets~\cite{rakthanmanonfast}, perform much worse in terms of accuracy compared to the state of the art~\cite{bagnall2016great}. As concrete example, consider the (actually rather small) PLAID benchmark dataset~\cite{gao2014plaid}, consisting of $1074$ profiles of $501$ measurements each stemming from $11$ different devices. Figure~\ref{fig:performance_on_plaid} plots classification times (in log scale) versus accuracy for seven state-of-the-art TSC methods and the novel algorithm presented in this paper, WEASEL. Euclidean distance (ED) based methods are the fastest, but their accuracy is far below standard. Dynamic Time Warping methods (DTW, DTW CV) are common baselines and show a moderate runtime of $10$ to $100$ ms but also low accuracy. Highly accurate classifiers such as ST~\cite{bostrom2015binary} and BOSS~\cite{schafer2014boss} require orders-of-magnitude longer prediction times. For this rather small dataset, the COTE ensemble classifier has not yet terminated training after right CPU weeks (Linux user time), thus we cannot report the accuracy, yet. In summary, the fastest methods for this dataset require around $1$ms per prediction, but have an accuracy below 80\%; the most accurate methods achieve 85\%-88\% accuracy, but require $80$ms up to $32$sec for each TS. 

In this paper, we propose a new TSC method called \emph{WEASEL: Word ExtrAction for time SEries cLassification}. WEASEL is both very fast and very accurate; for instance, on the dataset shown in Figure~\ref{fig:performance_on_plaid} it achieves the highest accuracy while being the third-fastest algorithm (requiring only $4$ms per TS). Like several other methods, WEASEL conceptually builds on the so-called bag-of-patterns approach: It moves a sliding window over a TS and extracts discrete features per window which are subsequently fed into a machine learning classifier. However, the concrete way of constructing and filtering features in WEASEL is completely different from any previous method. First, WEASEL considers differences between classes already during feature discretization instead of relying on fixed, data-independent intervals; this leads to a highly discriminative feature set. Second, WEASEL uses windows of varying lengths and also considers the order of windows instead of considering each fixed-length window as independent feature; this allows WEASEL to better capture the characteristics of each classes. Third, WEASEL applies aggressive statistical feature selection instead of simply using all features for classification; this leads to a much smaller feature space and heavily reduced runtime without impacting accuracy. The resulting feature set is highly discriminative, which allows us to use fast logistic regression instead of more elaborated, but also more runtime-intensive methods.

We performed a series of experiments to assess the impact of (each of) these improvements. First, we evaluated WEASEL on the popular UCR benchmark set of $85$ TS collections~\cite{UCRClassification} covering a variety of applications, including motion tracking, ECG signals, chemical spectrograms, and starlight-curves. WEASEL outperforms the best core-classifiers in terms of accuracy while also being one of the fastest methods; it is almost as accurate as the current overall best method (COTE) but multiple orders-of-magnitude faster in training and in classification. Second, for the concrete use case of energy load forecasting, we applied WEASEL to two real-live datasets and compared its performance to the other general TSC methods and to algorithms specifically developed and tuned for this problem. WEASEL again outperforms all other TS core-classifiers in terms of accuracy while being very fast, and achieves an accuracy on-par with the domain-specific methods without any domain adaptation. 

The rest of this paper is organized as follows: In Section~2 we present related work. Section~3 briefly recaps bag-of-patterns classifiers and feature discretization using Fourier transform. In Section~4 we present WEASEL's novel way of feature generation and selection. Section~5 presents evaluation results. The paper concludes with Section~6.

\section{Related Work}\label{sec:Background-and-Related}

With time series classification (TSC) we denote the problem of assigning a given TS to one of a predefined set of classes. TSC has applications in many domains; for instance, it is applied to determine the species of a flying insect based on the acoustic profile generated from its wing-beat~\cite{PotamitisSchaefer2014}, or for identifying the most popular TV shows from smart meter data~\cite{greveler2012multimedia}. 

The techniques used for TSC can be broadly categorized into two classes: whole series-based methods and feature-based methods~\cite{0001KL12}. \emph{Whole series} similarity measures make use of a point-wise comparison of entire TS. These include 1-NN Euclidean Distance (ED) or 1-NN Dynamic Time Warping (DTW)~\cite{rakthanmanon2012searching}, which is commonly used as a baseline in comparisons~\cite{lines2014time,bagnall2016great}. Typically, these techniques work well for short but fail for noisy or long TS~\cite{schafer2014boss}. Furthermore, DTW has a computational complexity of $O(n^{2})$ for TS of length $n$. Techniques like early pruning of candidate TS with cascading lower bounds can be applied to reduce the effective runtime~\cite{rakthanmanon2012searching}. Another speed-up techniques first clusters the input TS based on the fast ED and later analyzes the clusters using the triangle inequality~\cite{neamtu2016}.

In contrast, \emph{feature-based} classifiers rely on comparing features generated from substructures of TS. The most successful approaches can be grouped as either using shapelets or bag-of-patterns (BOP). \emph{Shapelets} are defined as TS subsequences that are maximally representative of a class. In~\cite{MueenKY11} a decision tree is built on the distance to a set of shapelets. The Shapelet Transform (ST)~\cite{lines2012shapelet,bostrom2015binary}, which is the most accurate shapelet approach according to a recent evaluation~\cite{bagnall2016great}, uses the distance to the shapelets as input features for an ensemble of different classification methods. In the Learning Shapelets (LS) approach~\cite{grabocka2014learning}, optimal shapelets are synthetically generated. The drawback of shapelet methods is the high computational complexity resulting in rather long training and classification times. 

The alternative approach within the class of feature-based classifiers is the \emph{bag-of-patterns (BOP)} model~\cite{0001KL12}. Such methods break up a TS into a bag of substructures, represent these substructures as discrete features, and finally build a histogram of feature counts as basis for classification. The first published BOP model (which we abbreviate as BOP-SAX) uses sliding windows of fixed lengths and transforms these measurements in each window into discrete features using Symbolic Aggregate approXimation (SAX)~\cite{Lin2007}. Classification is implemented as 1-NN classifier using Euclidean distance of feature counts as distance measure. SAX-VSM~\cite{senin2013sax} extends BOP-SAX with \emph{tf-idf} weighing of features and uses the Cosine distance; furthermore, it builds only one feature vector per class instead of one vector per sample, which drastically reduces runtime. Another current BOP algorithm is the TS bag-of-features framework (TSBF)~\cite{baydogan2013bag}, which first extracts windows at random positions with random lengths and next builds a supervised codebook generated from a random forest classifier. In our prior work, we presented the BOP-based algorithm BOSS (Bag-of-SFA-Symbols)~\cite{schafer2014boss}, which uses the Symbolic Fourier Approximation (SFA)~\cite{SchaferH12} instead of SAX. In contrast to shapelet-based approaches, BOP-based methods typically have only linear computational complexity for classification. 

The most accurate current TSC algorithms are Ensembles. These classify a TSC by a set of different core classifiers and then aggregate the results using techniques like bagging or majority voting. The Elastic Ensemble (EE PROP) classifier~\cite{lines2014time} uses $11$ whole series classifiers including DTW CV, DTW, LCSS and ED. The COTE ensemble~\cite{bagnalltime} is based on $35$ core-TSC methods including EE PROP and ST. If designed properly, ensembles combine the advantages of their core classifiers, which often lead to superior results. However, the price to pay is excessive runtime requirement for training and for classification, as each core classifier is used independently of all others.

\section{Time Series, BOP, and SFA}\label{sec:Symbolic-Representation-SFA}

The method we introduce in this paper follows the BOP approach and uses truncated Fourier transformations as first step on feature generation. In this section we present these fundamental techniques, after formally introducing time series and time series classification.

\begin{figure}
\begin{centering}
\includegraphics[width=1\columnwidth]{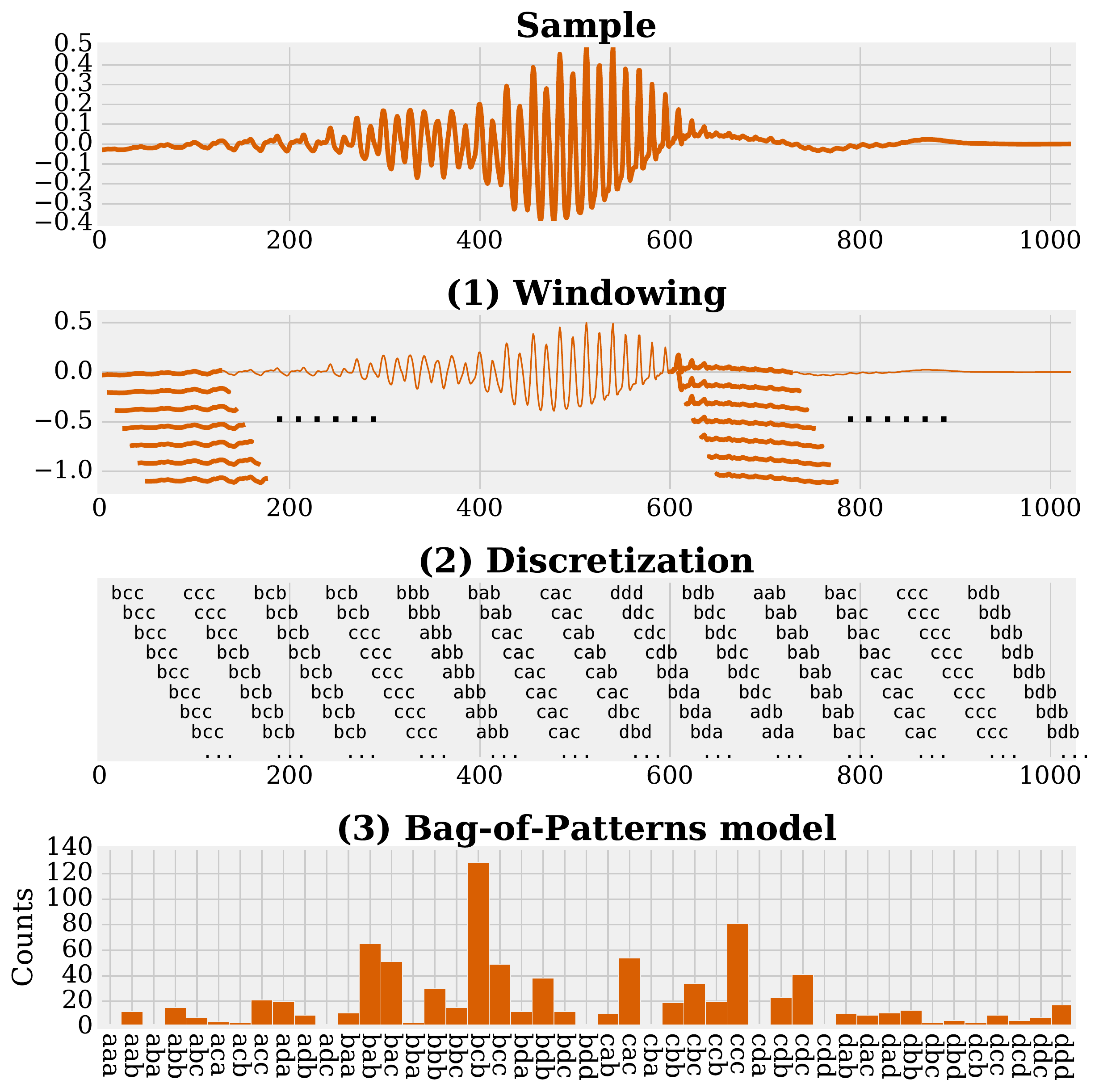}
\par\end{centering}
\caption{Transformation of a TS into the Bag-of-Patterns (BOP) model using overlapping windows (second to top), discretization of windows to words (second from bottom), and word counts (bottom).\label{fig:transformation}}
\end{figure}

In this work, a \emph{time series (TS)} $T$ is a sequence of $n\in\mathbb{N}$ real values, $T=(t_{1},\ldots,t_{n}),\;t_{i}\in\mathbb{R}$\footnote{Extensions to multivariate time series are discussed in Section 6}. As we primarily address time series generated from automatic sensors with a fixed sampling rate, we ignore time stamps. Given a TS $T$, a \emph{window} $S$ of length $w$ is a subsequence with $w$ contiguous values starting at offset $a$ in $T$, i.e., $S(a,w)=(t_{a},\text{\dots},t_{a+w-1})$ with $1\leq a\leq n-w+1$. We associate each TS with a class label $y\in Y$ from a predefined set $Y$. \emph{Time series classification (TSC)} is the task of predicting a class label for a TS whose label is unknown. A TS classifier is a function that is learned from a set of labeled time series (the training data), takes an unlabeled time series as input and outputs a label. 

Algorithms following the BOP model build this classification function by (1) extracting windows from a TS, (2) transforming each window of real values into a discrete-valued \emph{word} (a sequence of symbols over a fixed alphabet), (3) building a feature vector from word counts, and (4) finally using a classification method from the machine learning repertoire on these feature vectors. Figure~\ref{fig:transformation}  illustrates these steps from a raw time series to a BOP model using overlapping windows.

\begin{figure}
\begin{centering}
\includegraphics[width=1\columnwidth]{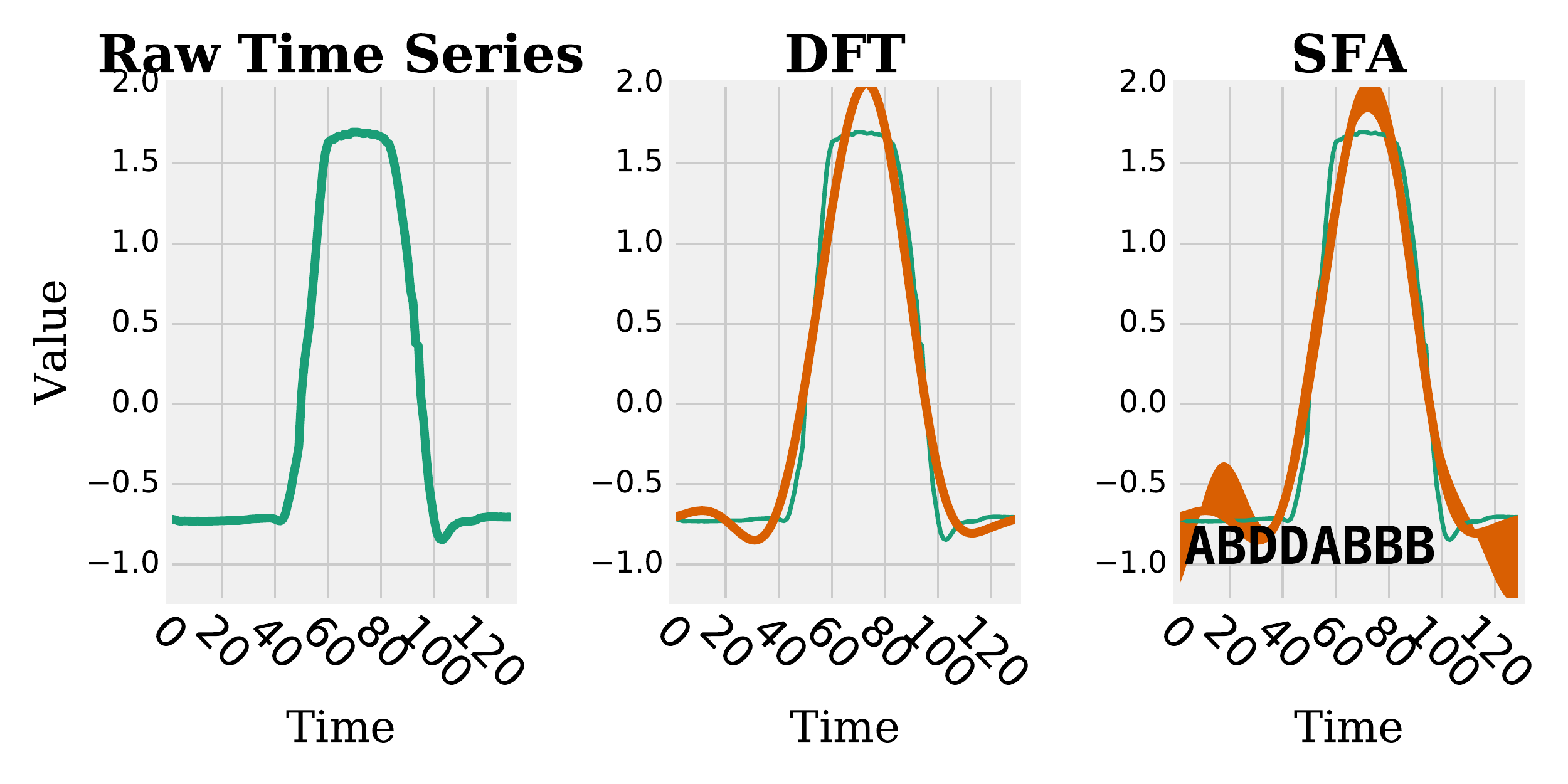}
\par\end{centering}
\caption{The Symbolic Fourier Approximation (SFA): A time series~(left) is approximated using the truncated Fourier transformation~(center) and discretized to the word \emph{ABDDABBB}~(right) with the four-letter alphabet ('a' to 'd'). The inverse transform is depicted by an orange area (right), representing the tolerance for all signals that will be mapped to the same word.\label{fig:SFATransform}}
\end{figure}

BOP methods differ in the concrete way of transforming a window of real-valued measurements into discrete words (discretization). WEASEL builds upon SFA which works as follows~\cite{SchaferH12}: (1) Values in each window are normalized to have standard deviation of $1$ to obtain amplitude invariance. (2) Each normalized window of length $w$ is subjected to dimensionality reduction by the use of the truncated Fourier Transform, keeping only the first $l<w$ coefficients for further analysis. This step acts as a low pass filter, as higher order Fourier coefficients typically represent rapid changes like dropouts or noise. (3) Each coefficient is discretized to a symbol of an alphabet of fixed size $c$ to achieve further robustness against noise. Figure~\ref{fig:SFATransform} exemplifies this process. 

\section{WEASEL}\label{sec:TSCWEASEL}

In this section, we present our novel TSC method WEASEL (Word ExtrAction for time SEries cLassification). WEASEL specifically addresses the major challenges any TSC method has to cope with when being applied to data from sensor readouts, which can be summarized as follows (using home device classification as an example): 

\emph{Invariance to noise}: TS can be distorted by (ambiance) noise as part of the recording process. In a smart grid, such distortions are created by imprecise sensors, information loss during transmission, stochastic differences in energy consumption, or interference of different consumers connected to the same power line. Identifying TS class-characteristic patterns requires to be noise robust.

\emph{Scalability}: TS in sensor-based applications are typically recorded with high sampling rates, leading to long TS. Furthermore, smart grid applications typically have to deal with thousands or millions of TS. TSC methods in such areas need to be scalable in the number and length of TS.

\emph{Variable lengths and offsets}: TS to be classified may have variable lengths, and recordings of to-be-classified intervals can start at any given point in time. In a smart grid, sensors produce continuous measurements, and the partitioning of this essentially infinite stream into classification intervals is independent from the usages of devices. Thus, characteristic patterns may appear anywhere in a TS (or not at all), but typically in the same order.

\emph{Unknown characteristic substructures}: Feature-based classifiers exploit local substructures within a TS, and thus depend on the identification of recurring, characteristic patterns. However, the position, form, and frequency of these patterns is unknown; many substructures may be irrelevant for classification. For instance, the idle periods of the devices in Figure~\ref{fig:DailyPowerConsumption} are essentially identical.

We carefully engineered WEASEL to address these challenges. Our method conceptually builds on the BOP model in BOSS~\cite{schafer2014boss}, yet uses rather different approaches in many of the individual steps. We will use the terms \emph{feature} and \emph{word} interchangeably throughout the text. Compared to previous works in TSC, WEASEL implements the following novel ideas, which will be explained in detail in the following subsections:

\begin{enumerate}

\item \textbf{Discriminative feature generation}: WEASEL derives discriminative features based on class characteristics of the concrete dataset.
This differs from current BOP~\cite{Lin2007,SchaferH12} methods, which apply the same feature generation method independent of the actual dataset, possibly leading to features that are equally frequent in all classes, and thus not discriminative. 
Specifically, our approach first Fourier transforms each window, next determines discriminative Fourier coefficients using the ANOVA f-test and finally applies information gain binning for choosing appropriate discretization boundaries. Each step aims at separating TS from different classes. 

\item \textbf{Co-occurring words}: The order of substructures (each represented by a word) is lost in the BOP model. To mitigate this effect, WEASEL also considers bi-grams of words as features. Thus, local order is encoded into the model, but as a side effect the feature space is increased drastically. 

\item \textbf{Variable-length windows}: Typically, characteristic TS patterns do not all have the same length. Current BOP approaches, however, assume a fixed window length, which leads to ignorance regarding patterns of different lengths. WEASEL removes this restriction by extracting words for multiple window lengths and joining all resulting words in a single feature vector - instead of training separate vectors and selecting (the best) one as in  other BOP models. This approach can capture more relevant signals, but again increases the feature space. 

\item \textbf{Feature selection}: The wide range of features considered captures more of the characteristic TS patterns but also introduces many irrelevant features. Therefore, WEASEL uses an aggressive Chi-Squared test to filter the most relevant features in each class and reduce the feature space without negatively impacting classification accuracy. 
\end{enumerate}

\begin{figure}
\begin{centering}
\includegraphics[width=1\columnwidth]{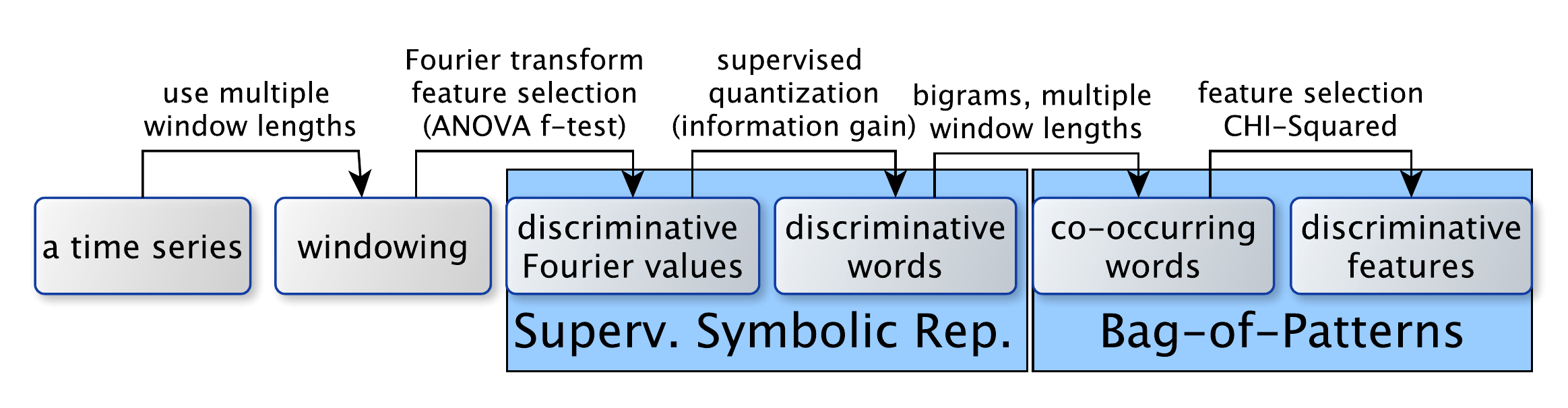}
\par\end{centering}
\caption{WEASEL Pipeline: Feature extraction using our novel supervised symbolic representation, the novel bag-of-patterns model, and feature matching using a logistic regression classifier.\label{fig:WEASEL-Pipeline}}
\end{figure}

WEASEL is composed of the building blocks depicted in Figure~\ref{fig:WEASEL-Pipeline}: our novel supervised symbolic representation for discriminative feature generation and the novel bag-of-patterns model for building a discriminative feature vector. 
First, WEASEL extracts normalized windows of different lengths from a time series. Next, each window is approximated using the Fourier transform, and those Fourier coefficients are kept that best separate TS from different classes using the ANOVA F-test. The remaining Fourier coefficients are discretized into a word using information gain binning, which also chooses discretization boundaries to best separate the TS classes; More detail is given in Subsection~4.2.
Finally, a single bag-of-patterns is built from the words (unigrams) and neighboring words (bigrams). This bag-of-patterns encodes unigrams, bigrams and windows of variable lengths. To filter irrelevant words, the Chi-Squared test is applied to this bag-of-patterns (Subsection~4.1). As WEASEL builds a highly discriminative feature vector, a fast linear time logistic regression classifier is applied, as opposed to more complex, quadratic time classifiers (Subsection~4.1).

\begin{algorithm}[t]
{\scriptsize{}}
\begin{lstlisting}[language=Java,numbers=left,basicstyle={\scriptsize\sffamily},breaklines=true,showstringspaces=false,tabsize=2,backgroundcolor={\color{gray}},numbersep=1em,xleftmargin=2em,xrightmargin=0em,emph={function, in, all, each, to},emphstyle={\textbf},escapechar={Ã}]
function Ã\textbf{WEASEL}Ã(sample, l)
	bag = empty BagOfPattern

	// extract words for each window length
	for each window length w:
		allWindows = Ã\texttt{SLIDING\_WINDOWS}Ã(sample, w)
		Ã\texttt{norm}Ã(allWindows)

		for each (prevWindow, window) in allWindows:
			// BOP computed from unigrams
			word = Ã\texttt{quantization.transform}Ã(window,l)
			bag[w++word].increaseCount()

			// BOP computed from bigrams
			prevWord=Ã\texttt{quantization.transform}Ã(prevWindow,l)	
			bag[w++prevWord++word].increaseCount()

	// feature selection using ChiSquared
	return Ã\texttt{CHI\_SQUARED\_FILTERED}Ã(bag)
\end{lstlisting}
{\scriptsize \par}

\caption{Build one bag-of-patterns using a supervised symbolic representation,
multiple window lengths, bigrams and the Chi-squared test for feature
selection. $l$ is the number of Fourier values to keep.\label{alg:The-WEASEL-representation}}
\end{algorithm}

Algorithm~\ref{alg:The-WEASEL-representation} illustrates WEASEL: sliding windows of length $w$ are extracted (line~6) and windows are normalized (line~7). We empirically set the window lengths to all values in $[8,\dots,n]$. Smaller values are possible, but the feature space can become untraceable, and small window lengths are basically meaningless for TS of length $>10^3$.
Our supervised symbolic transformation is applied to each real-valued sliding window (line~11,15). Each word is concatenated with the window length and its occurrence is counted (line~12,16). Lines~15\textendash16 illustrate the use of bigrams: the preceding sliding window is concatenated with the current window. Note, that all words (unigrams,bigrams,window-length) are joined within a single bag-of-patterns. Finally irrelevant words are removed from this bag-of-patterns using the Chi-Squared test~(line~19). The target dimensionality $l$ is learned through cross-validation.

\begin{figure}[t]
\includegraphics[width=1\columnwidth]{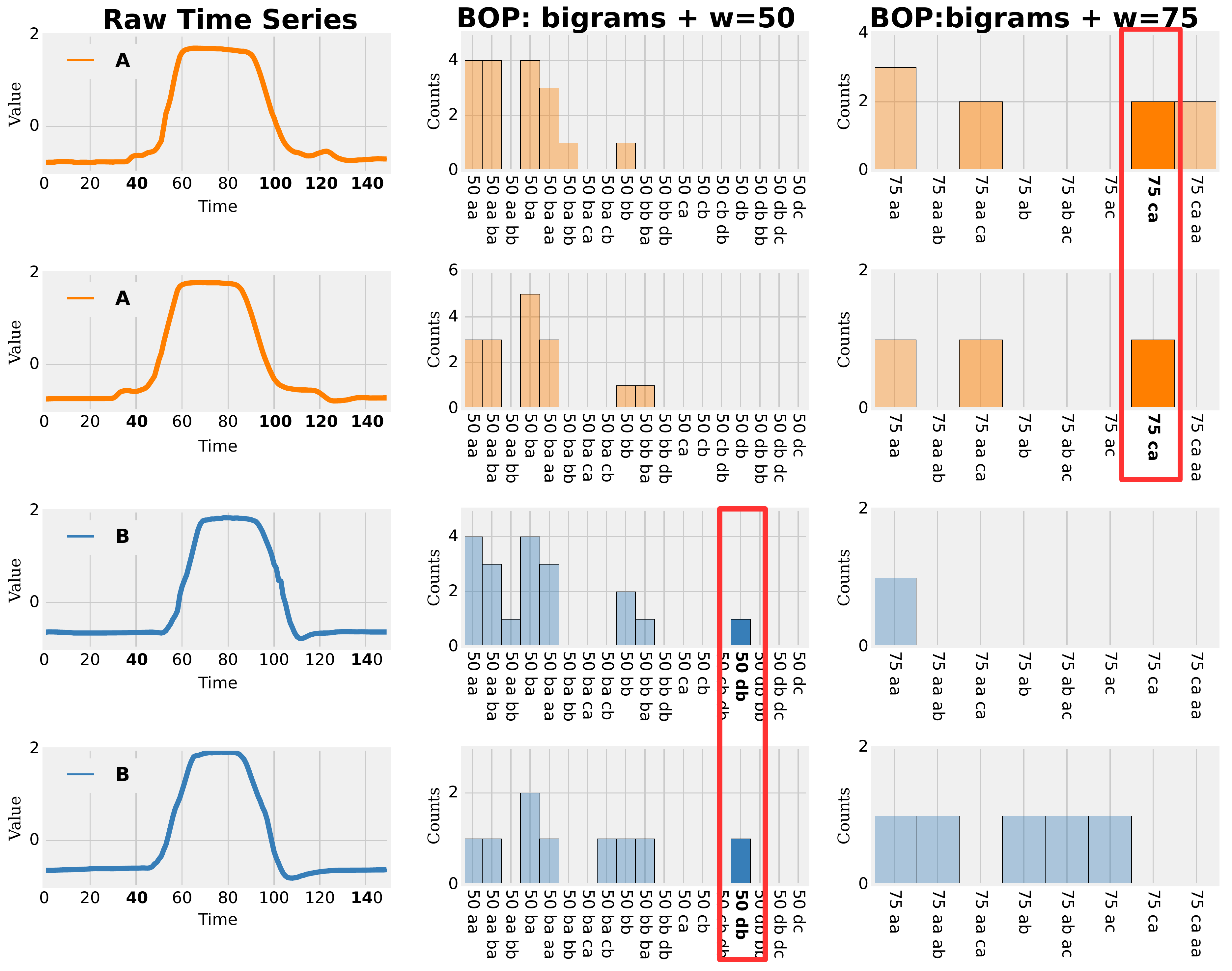}


\caption{Discriminative feature vector: Four time series, two from class 'A' and two from class 'B' are shown. Feature vectors contain unigrams and bigrams for the window lengths $w$ of $50$ and $75$. The discriminative words are highlighted.\label{fig:Bigrams-Logistic}}

\end{figure}

BOP-based methods have a number of parameters, which heavily influence their performance. Of particular importance is the window  length $w$. An optimal value for this parameter is typically learned for each new dataset using techniques like cross-validation. This does not only carry the danger of over-fitting (if the training samples are biased compared to the to-be-classified TS), but also leads to substantial training times. In contrast, WEASEL removes the need to set this parameter, by constructing one joined high-dimensional feature vector, in which every feature encodes the parameter's values (Algorithm~\ref{alg:The-WEASEL-representation} lines~12,16). 

Figure~\ref{fig:Bigrams-Logistic} illustrates our use of unigrams, bigrams and variable window lengths. The depicted dataset contains two classes 'A' and 'B' with two samples each. The time series are very similar and differences between these are difficult to spot, and are mostly located between time stamps $80$ and $100$ to $130$. The center (right) column illustrates the features extracted for window length $50$ ($75$).
Feature \emph{'75 aa ca'} (a bigram for length 75) is characteristic for the \emph{A} class, whereas the feature \emph{'50 db'} (an unigram for length 50) is characteristic for the \emph{B} class. Thus, we use different window lengths, bigrams, and unigrams to capture subtle differences between TS classes. 
We show the impact of variable-length windows and bigrams to classification accuracy in Section~\ref{influence}.

\subsection{Feature Selection and Weighting: Chi-squared Test and Logistic Regression}\label{subsec:Chi-Squared-Test}

The dimensionality of the BOP feature space is $O(c^{l})$ for word length $l$ and $c$  symbols. It is independent of the number of time series $N$ as these only affect the frequencies. For common parameters like $c=4$, $l=4$, $n=256$ this results in a sparse vector with $4^{4} = 256$ dimensions for a TS. WEASEL uses bigrams and $O(n)$ window lengths, thus the dimensionality of the feature space rises to $O(c^{l}\cdot~c^{l}\cdot~n)$. For the previous set of parameters this feature space explodes to $4^{8}\cdot~256 = 256^{3}$. 

WEASEL uses the Chi-squared ($\chi^{2}$) test to identify the most relevant features in each class to reduce this feature space to a few hundred features prior to training the classifier. This statistical test determines if for any feature the observed frequency within a specific group significantly differs from the expected frequency, assuming the data is nominal. Larger $\chi^{2}$-values imply that a feature occurs more frequently within a specific class. Thus, we keep those features with $\chi^{2}$-values above the threshold. This highlights subtle distinctions between classes. All other features can be considered superfluous and are removed.  On average this reduces the size of the feature space by $30-70\%$ to roughly $10^4$ to $10^5$ features.

Still, with thousands of time series or features an accurate, quadratic time classifier can take days to weeks to train on medium-sized datasets~\cite{schafer2015scalable2}. For sparse vectors, linear classifiers are among the fastest, and they are known to work well for large dimensional (sparse) vectors, like in document
classification. These linear classifiers predict the label based on a dot-product of the input feature vector and a weight vector. The weight vector represents the model trained on labeled train samples. Using a weight vector highlights features that are characteristic for a class label and suppresses irrelevant features. Thus, the classifier aims at finding those features, that can be used to determine a class label. Methods to obtain a weight vector include Support Vector Machines~\cite{cortes1995support} or logistic regression~\cite{freedman2009statistical}. We implemented our classifier using liblinear~\cite{fan2008liblinear} as it scales linearly with the dimensionality of the feature space~\cite{ng2004feature}. This results in a moderate runtime compared to Shapelet or ensemble classifiers, which can be orders of magnitude slower (see Section 5.3). 

\subsection{Supervised Symbolic Representation}\label{subsec:SupervisedSymbolicRepresentation}

A symbolic representation is needed to transform a real-valued TS window to a word using an alphabet of size $c$. The problem with SFA~\cite{SchaferH12} is that it (a) filters the high frequency components of the signal, just like a low-pass filter. But for instance, the pitch (frequency) of a bird sound is relevant for the species but lost after low-pass filtering. Furthermore, it (b) does not distinguish between class labels when quantizing values of the Fourier transform. Thus, there is a high likelihood of SFA words to occur in different classes with roughly equal frequencies. For classification, we need discriminative words for each class. Our approach is based on:
\begin{enumerate}
\item \textbf{Discriminative approximation}: We introduce feature selection to the approximation step by using the one-way ANOVA F-test: we keep the Fourier values whose distribution best separates the class labels in disjoint groups.
\item \textbf{Discriminative quantization}: We propose the use of information gain~\cite{quinlan1986induction}. This minimizes the entropy of the class labels for each split. I.e., the majority of values in each partition correspond to the same class label. 
\end{enumerate}

\begin{figure}[t]
\includegraphics[width=1\columnwidth]{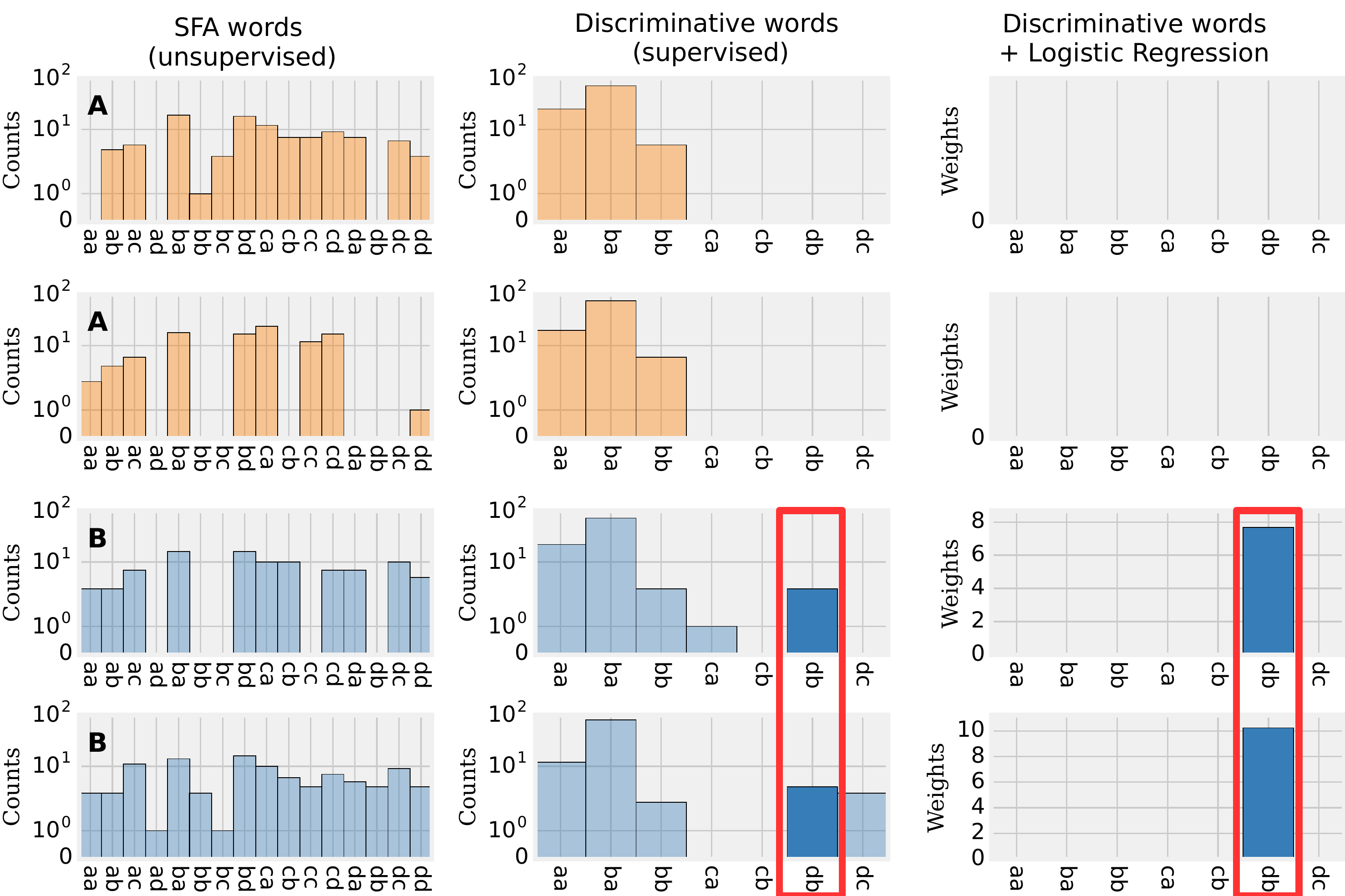}

\caption{Influence of the word model for feature extraction. From left to right: bag-of-patterns for SFA words, our novel discriminative words, and our weighted discriminative words after logistic regression. \label{fig:Word-Histograms}}

\end{figure}

In Figure~\ref{fig:Word-Histograms} we revisit our sample dataset. This time with a window length of $25$. When using SFA words (left), the words are evenly spread over the whole bag-of-patterns for both prototypes. There is no single feature whose absence or presence is characteristic for a class. 
However, when using our novel discriminative words (center), we observe less distinct words, more frequent counts and the word \emph{'db'} is unique within the \emph{'B'} class. Thus, any subsequent classifier can separate classes just by the occurrence of this feature. When training a logistic regression classifier on these words (right), the word \emph{'db'} gets boosted and other words are filtered.
Note, that the counts of the word \emph{'db'} differ for both representations, as it represents other frequency ranges for the SFA and discriminative words.
This showcase underlines that not only different window lengths or bigrams (as in Figure~\ref{fig:Bigrams-Logistic}), but also the symbolic representation helps to generate discriminative feature sets. Our showcase is the Gun-Point dataset~\cite{UCRClassification}, which represents the hand movement of actors, who aim a gun (prototype A) or point a finger (prototype B) at people.

\subsubsection{Discriminative Approximation using One-Way ANOVA F-test} \

\begin{figure}[t]
\includegraphics[width=1\columnwidth]{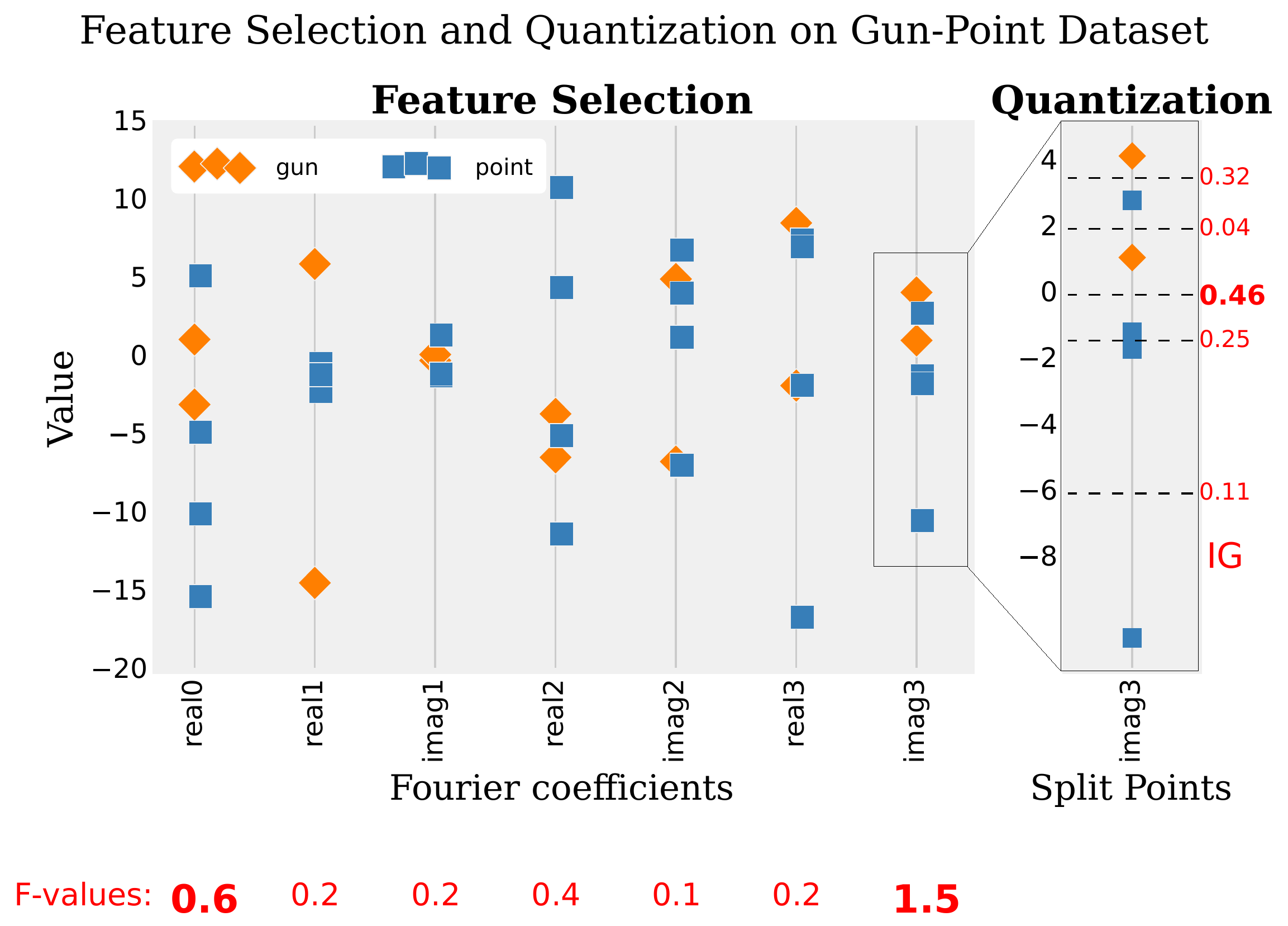}

\caption{On the left: Distribution of Fourier coefficients for the samples from the Gun-Point dataset. The high F-values on imag3 and real0 (red text at bottom) should be selected to best separate the samples from class labels 'Gun' and 'Point'. On the right: Zoom in on imag3. Information gain splitting is applied to find the best bins for the subsequent quantization. High information gain (IG) indicates pure (good) split points.
\label{fig:ANOVA-F-Test}}
\end{figure}

For approximation each TS is Fourier transformed first. We aim at finding those real an imaginary Fourier values that best separate between class labels for a set of TS samples, instead of simply taking the first ones. Figure~\ref{fig:ANOVA-F-Test} (left) shows the distribution of the Fourier values for the samples from the Gun-Point dataset. The Fourier value that best separates between the classes is $\textit{imag}_{3}$ with the highest F-value of $1.5$ (bottom).

We chose to use a one-way ANOVA F-test~\cite{lowry2014concepts} to select the best Fourier coefficients, as it is applicable on continuous variables, as opposed to the Chi-squared test, which is limited to categorical variables. The one-way ANOVA F-test checks the hypothesis that two or more groups have the same normal distribution around the mean. The analysis is based on two estimates for the variance existing within and between groups: \emph{mean square within} ($MS_{W}$) and \emph{mean square between} ($MS_{B}$). The F-value is then defined as: $F=\frac{MS_{B}}{MS_{W}}$. If there is no difference between the group means, the F-value is close to or below $1$. If the groups have different distributions around the mean, $MS_{B}$ will be larger than $MS_{W}$. When used as part of feature selection, we are interested in the largest F-values, equal to large differences between group means. The F-value is calculated for each real $real_{i}\in REAL(T)$ and imaginary $imag_{i}\in IMAG(T)$ Fourier value. We keep those $l$ Fourier values with the largest F-values. In Figure~\ref{fig:ANOVA-F-Test} these are $\textit{real}_{0}$ and $\textit{imag}_{3}$ for $l=2$ with F-values $0.6$ and $1.5$.

Assumptions made for the ANOVA F-test:

\begin{enumerate}

\item The ANOVA F-test assumes that the data follows a normal distribution with equal variance. The BOP (WEASEL) approach extracts subsequences for z-normalized time series. It has been shown that subsequences extracted from z-normalized time series perfectly mimic normal distribution~\cite{Lin2003}. Furthermore, the Fourier transform of a normal distribution \[
f(x)=\frac{1}{\sigma\sqrt{2\pi}}\cdot e^{-\frac{-x^{2}}{2\sigma^{2}}}
\]

with $\mu=0,\sigma=1$ results in a normal distribution of the Fourier coefficients~\cite{bryc2012normal}: \[
F(t)=\int f(x)\cdot e^{-itx}=e^{i\mu\sigma}e^{-\frac{1}{2}(\sigma t)^{2}}=e^{-\frac{1}{2}(\sigma t)^{2}}
\]
Thus, the Fourier coefficients follow a symmetrical and uni-modal normal distribution with equal variance.

\item The ANOVA F-test assumes that the samples are independently drawn. To guarantee independence, we are extracting disjoint subsequences, i.e. non-overlapping, to train the quantization intervals. Using disjoint windows for sampling further decreases the likelihood of over-fitting quantization intervals.

\end{enumerate}

\subsubsection{Discriminative Quantization using Entropy / Information Gain} \

A quantization step is applied to find for each selected real or imaginary Fourier value the best split points, so that in each partition a majority of values correspond to the same class. We use information gain~\cite{quinlan1986induction} and search for the split with largest information gain, which represents an increase in purity. Figure~\ref{fig:ANOVA-F-Test} (right) illustrates five possible split points for the $imag_{3}$ Fourier coefficient on the two labels 'Gun' (orange) and 'Point' (red). The split point with the highest information gain of $0.46$ is chosen.

Our quantization is based on binning (bucketing). The value range is partitioned into disjoint intervals, called bins. Each bin is labeled by a symbol. A real value that falls into an interval is represented by its discrete label. Common methods to partition the value range include equi-depth or equi-width bins. These ignore the class label distribution and splits are solely based on the value distribution. Here we introduce entropy-based binning. This leads to disjoint feature sets. Let $Y=\left\{ (s_{1},y_{1}),\ldots,(s_{N},y_{N})\right\} $ be a list
of value and class label pairs with $N$ unique class labels. The multi-class entropy is then given by: $Ent(Y)=\sum_{(s_{i},y_{i})\in Y}-p_{y_{i}}\log_{2}p_{y_{i}}$, where $p_{y_{i}}$ is the relative frequency of label $y_{i}$ in $Y$. The entropy for a split point $sp$ with all labels on the left $Y_{L}=\left\{ (s_{i},y_{i})\left|s_{i}\leq sp,\,(s_{i},y_{i})\in Y\right.\right\} $ and all labels on the right $Y_{R}=\left\{ (s_{i},y_{i})\left|s_{i}>sp,\,(s_{i},y_{i})\in Y\right.\right\} $ is given by:
\begin{equation}
Ent(Y,sp)=\frac{\left|Y_{L}\right|}{\left|Y\right|}Ent(Y_{L})+\frac{\left|Y_{R}\right|}{\left|Y\right|}Ent(Y_{R})
\end{equation}

The information gain for this split is given by: 
\begin{equation}
\textit{Information Gain}=Ent(Y)-Ent(Y,sp)
\end{equation}

\begin{algorithm}[t]
{\scriptsize{}}
\begin{lstlisting}[language=Java,numbers=left,basicstyle={\scriptsize\sffamily},breaklines=true,showstringspaces=false,tabsize=2,backgroundcolor={\color{gray}},numbersep=1em,xleftmargin=2em,xrightmargin=0em,emph={function, in, all, each, to},emphstyle={\textbf},escapechar={Ã}]
function Ã\textbf{FitBins}Ã(dfts, l, c)
	bins[l][c] // 2d array of bins
	for i = 1 to l:
		// order line of class labels sorted by value
		orderLine = buildOrderLine(dfts, i) 		

		Ã\texttt{IGSplit}Ã(orderLine, bins[i])
	return bins

function Ã\textbf{IGSplit}Ã(orderLine, bins)
	(sp, Y_L, Y_R) = find split with maximal IG
	bins.add(sp)
	if (remaining bins): // recursive splitting
		Ã\texttt{IGSplit}Ã(Y_L, bins) // left
		Ã\texttt{IGSplit}Ã(Y_R, bins) // right
\end{lstlisting}
{\scriptsize \par}

\caption{Entropy-based binning of the real and imaginary Fourier values.\label{alg:Entropy-based-binning-of}}
\end{algorithm}

Algorithm~\ref{alg:Entropy-based-binning-of} illustrates entropy-binning for a $c$ symbol alphabet and word length $l$. For each set of the $l$ real and imaginary Fourier values, an order-line is built (line~5). We then search for the $c$ split points that maximize the information gain (line~6). After choosing the first split point (line~10) any remaining partition $Y_{L}$ or $Y_{R}$ that is not pure is recursively split (lines~13-14). The recursion ends once we have found $c$ bins (line~12). 

We fix the alphabet size c to $4$, as it has been shown in the context of BOP models that using a constant $c=4$ is very robust over all TS considered~\cite{0001KL12, senin2013sax, schafer2014boss}. 

\section{Evaluation}\label{sec:Results}
\subsection{Experimental Setup}

We mostly evaluated our \emph{WEASEL} classifier using the full UCR benchmark dataset of 85 TSC problems~\cite{UCRClassification}\footnote{The UCR archive has recently been extended from 45 to 85 datasets.}. Furthermore, we compared its performance on two real-life datasets from the smart grid domain; results are reported in Section~\ref{use_case}.

Each UCR dataset provides a train and test split set which we use unchanged to make our results comparable the prior publications. We compare  WEASEL to the best published TSC methods (following~\cite{bagnall2016great}), namely COTE (Ensemble)~\cite{bagnalltime}, 1-NN BOSS~\cite{schafer2014boss}, Learning Shapelets~\cite{grabocka2014learning}, Elastic Ensemble (EE PROP)~\cite{lines2014time}, Time Series Bag of Features (TSBF)~\cite{baydogan2013bag}, Shapelet Transform (ST)~\cite{bostrom2015binary}, and 1-NN DTW with and without a warping window set through cross validation on the training data (CV)~\cite{lines2014time}. A recent  study~\cite{bagnall2016great} reported COTE, ST, BOSS, and EE PROP as the most accurate (in this order).

All experiments ran on a server running LINUX with 2xIntel Xeon E5-2630v3 and 64GB RAM, using JAVA JDK x64 1.8. We measured runtimes of all methods using the implementation given by the authors~\cite{BOSSGitHub} wherever possible, resorting to the code by~\cite{bagnall2016great} if this was not the case. For 1-NN DTW and 1- NN DTW CV, we make use of the state-of-the-art cascading lower bounds from~\cite{rakthanmanon2012searching}. Multi-threaded code is available for BOSS and WEASEL, but we have restricted all codes to use a single core to ensure comparability of numbers. Regarding accuracy, we report numbers published by each author~\cite{bagnalltime,BagnallDHL12,grabocka2014learning,baydogan2013bag}, complemented by the numbers published by~\cite{TimeSeriesClassification}, for those datasets where results are missing (due to the growth of the benchmark datasets). All numbers are accuracy on the test split. 

For WEASEL we performed $10$-fold cross-validation on the training datasets to find the most appropriate value for the SFA word length $l\in\left[4,6,8\right]$ We kept $c=4$ and $chi=2$ constant, as varying these values has only negligible effect on accuracy (data not shown). We used \emph{liblinear} with default parameters $(\textit{bias}=1,p=0.1$ and solver L2R\_LR\_DUAL). To ensure reproducible results, we provide the WEASEL source code and the raw measurement sheets~\cite{WEASELWebPage}. 

\subsection{Accuracy}

\begin{figure}[t]
\begin{centering}
\includegraphics[width=1\columnwidth]{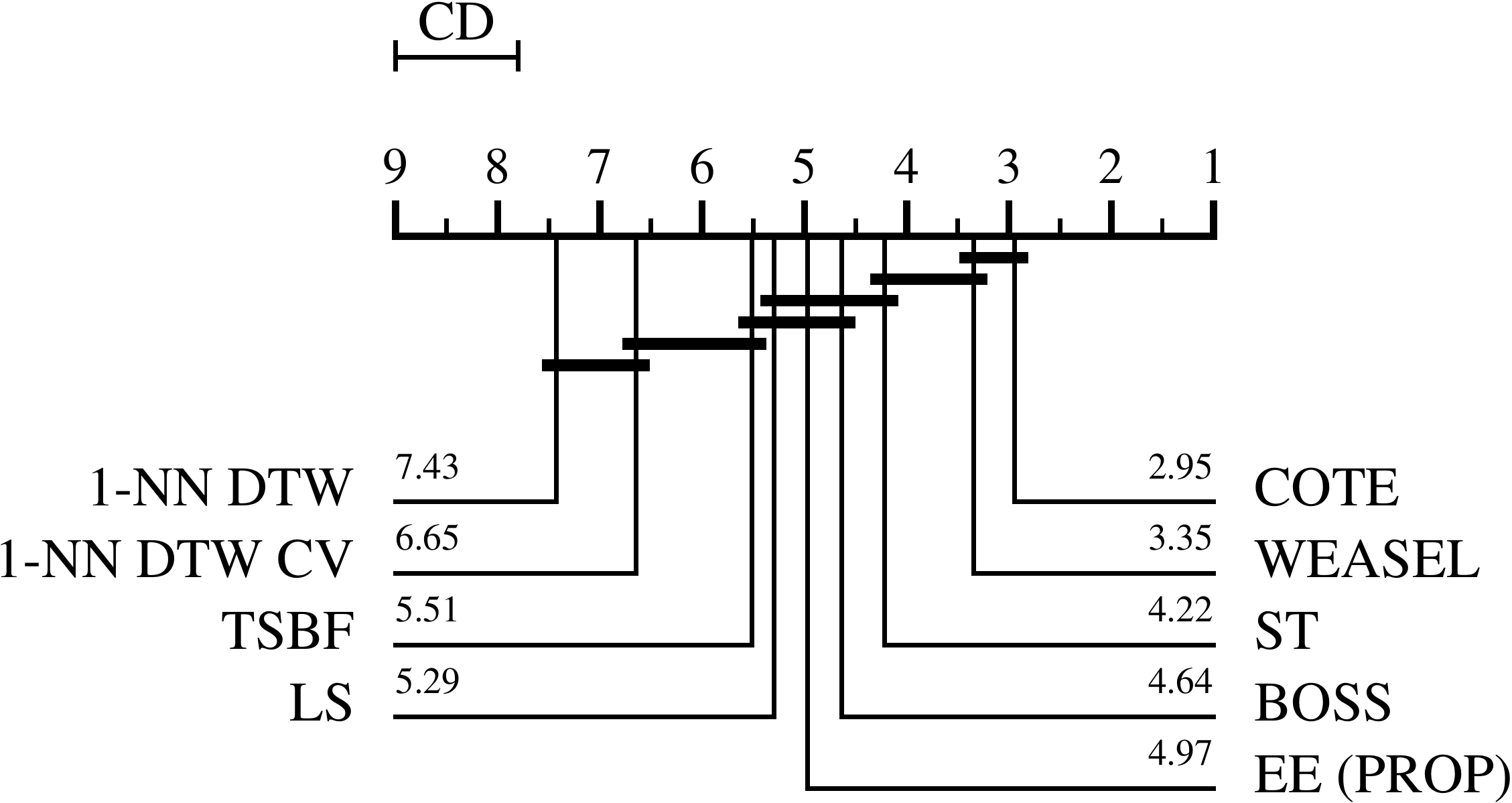}
\end{centering}
\caption{Critical difference diagram on average ranks on 85 benchmark datasets. WEASEL is as accurate as state of the art.\label{fig:Average-Ranks-on}}
\end{figure}

Figure~\ref{fig:Average-Ranks-on} shows a critical difference diagram (introduced in~\cite{demvsar2006statistical}) over the average ranks of  of the different TSC methods. Classifiers with the lowest (best) ranks are to the right. The group of classifiers that are not significantly different in their rankings are connected by a bar. The critical difference (CD) length, which represents statistically significant differences, is shown above the graph. 

The 1-NN DTW and 1-NN DTW CV classifiers are commonly used as benchmarks~\cite{lines2014time}. Both perform significantly worse than all other methods. Shapelet Transform (ST), Learning Shapelets (LS) and BOSS have a similar rank and competitive accuracies. WEASEL is the best (lowest rank among all core classifiers (DTW, TSBF, LS, BOSS, ST), i.e., it is on average the most accurate core classifiers. This confirms our assumptions that the WEASEL pipeline resembles the requirements for time series similarity (see Section~5.3).

Ensemble classifiers generally show compelling accuracies at the cost of enormous runtimes. The high accuracy is confirmed in Figure~\ref{fig:Average-Ranks-on}, where COTE~\cite{bagnalltime} is the overall best method. The advantage of WEASEL is its much lower runtime, which we address in Section~5.3.

We performed a Wilcoxon signed rank test to assess the differences between WEASEL and COTE, ST, BOSS, EE. The p-values are $0.0001$ for BOSS, $0.017$ for ST, $0.0000032$ for EE and COTE $0.57$.  Thus, at a cutoff of $p=0.05$, WEASEL is  significantly better than BOSS, ST and EE, yet very similar to COTE.

\begin{figure}[t]
\includegraphics[width=1\columnwidth]{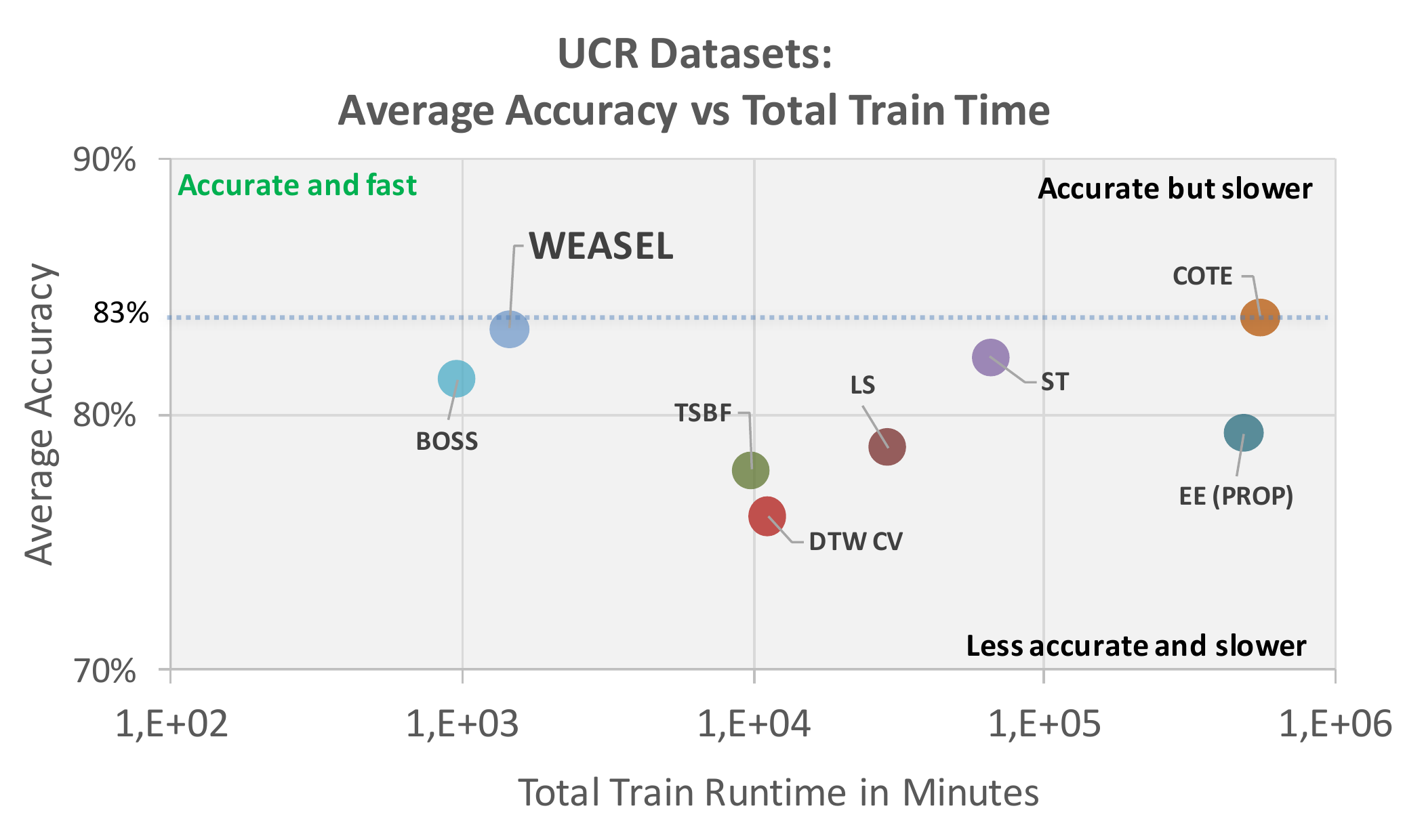}
\includegraphics[width=1\columnwidth]{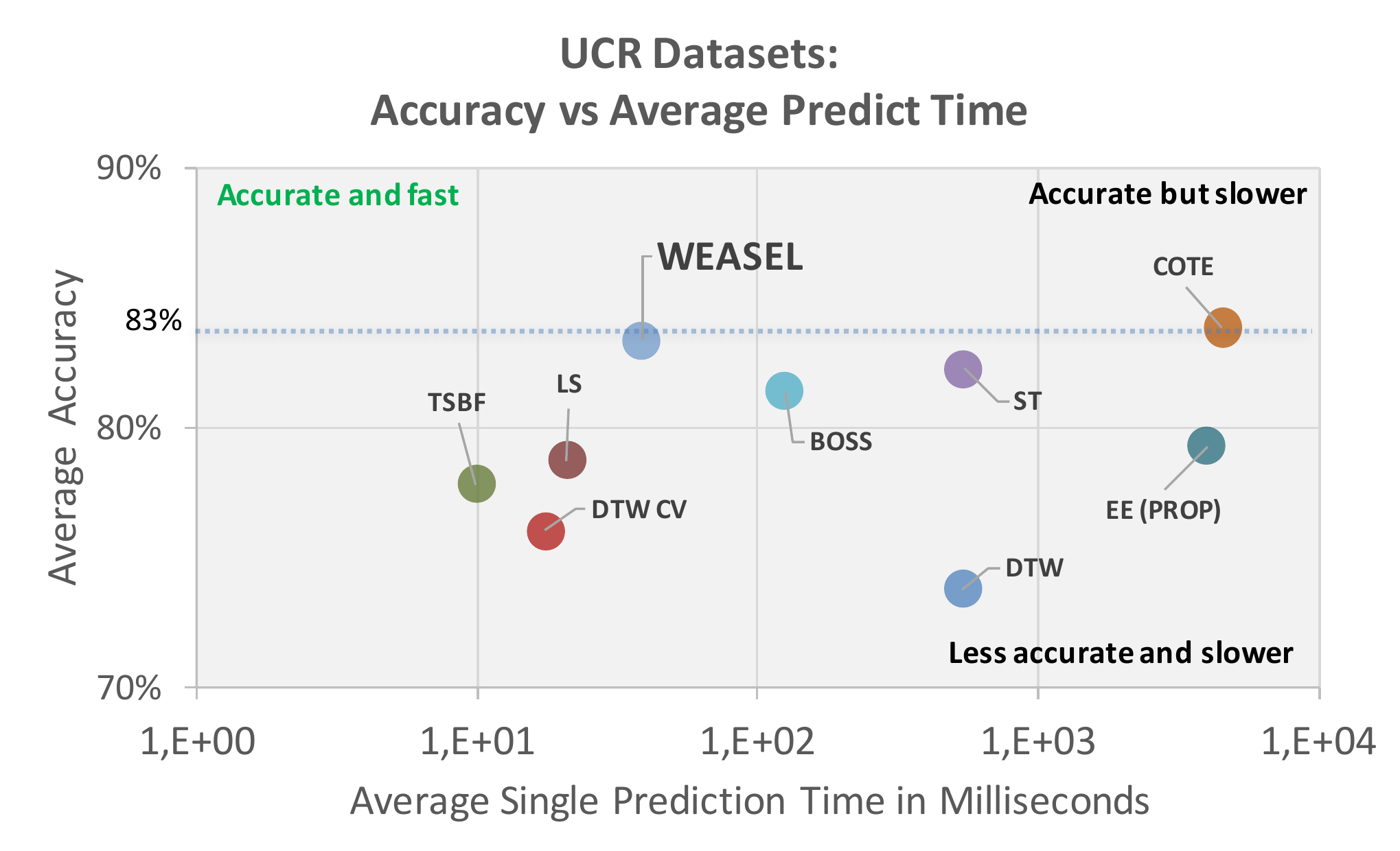}
\caption{Average single prediction (top) / total training time (bottom) in log scale vs average accuracy. Runtimes include all preprocessing steps (feature extraction, bop model building, etc.). WEASEL has a similar average accuracy as COTE but is two orders of magnitude faster. A single prediction takes $38$ms on average.\label{fig:Time-vs-Accuracy}}
\end{figure}

\vspace{2em}

\subsection{Scalability}
Figure~\ref{fig:Time-vs-Accuracy} plots for all TSC methods the total runtime on the x-axis in log scale vs the average accuracy on the y-axis for training (top) and prediction (bottom). Runtimes include all preprocessing steps like feature extraction or selection. Because of the high wall-clock time of some classifiers, we limited this experiment to the $45$ core UCR datasets, encompassing roughly $N=17000$ train and $N=62000$ test time series. The slowest classifiers took more than $340$ CPU days to train (Linux user time).

The DTW classifier is the only classifier that does not require training. The DTW CV classifier requires a training step to set a warping window, which significantly reduces the runtime for the prediction step. Training DTW CV took roughly $186$ CPU hours until completion. WEASEL and BOSS  have similar train times of $16-24$ CPU hours and are one to two orders of magnitude faster than the other core classifiers. WEASEL's prediction time is $38$ms on average and one order of magnitude faster than that of BOSS. LS and TSBF have the lowest prediction times but a limited average accuracy~\cite{bagnall2016great,schafer2015scalable2}. As expected, the two Ensemble methods in our comparison, EE PROP and COTE, show by far the longest training and classification times. On the NonInvasiveFatalECGThorax1, NonInvasiveFatalECGThorax2, and StarlightCurves datasets training each ensemble took more than $120, 120$ and $45$ CPU days. 



\begin{figure*}[t]
\includegraphics[width=2\columnwidth]{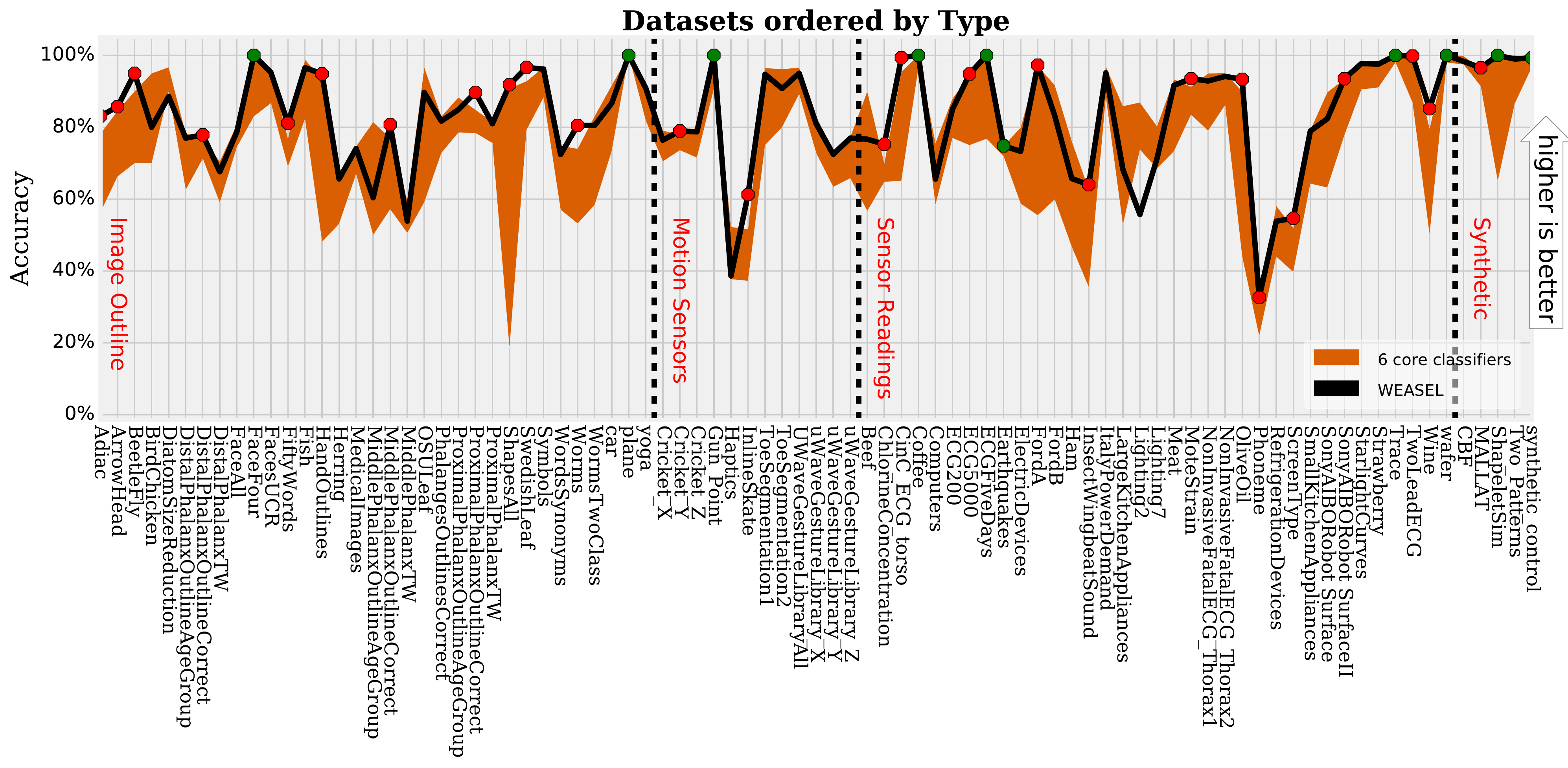}
\caption{Classification accuracies for WEASEL vs the best six core classifiers (ST, LS, BOSS, DTW, DTW CV, TSBF). The orange area represents the six core classifiers' accuracies. Red (green) dots indicate where WEASEL wins (evens out) against the other classifiers.\label{fig:Classification-accuracy-for}}
\end{figure*}

\vspace{3em}

\subsection{Accuracy by datasets and by domain}

\begin{table*}
\begin{centering}
\begin{tabular}{cccccccc}
\toprule 
 & \textbf{WEASEL} & \textbf{DTW CV} & \textbf{DTW} & \textbf{BOSS} & \textbf{LS} & \textbf{TSBF} & \textbf{ST}\tabularnewline
\midrule
\midrule 
Image Outline & 39.4\% & 12.1\% & 9.1\% & 27.3\% & 3.0\% & 18.2\% & 21.2\%\tabularnewline
\midrule 
Motion Sensors & 25.0\% & 8.3\% & 0.0\% & 16.7\% & 25.0\% & 25.0\% & 25.0\%\tabularnewline
\midrule 
Sensor Readings & 48.6\% & 8.6\% & 8.6\% & 17.1\% & 20.0\% & 14.3\% & 25.7\%\tabularnewline
\midrule 
Synthetic & 60.0\% & 0.0\% & 20.0\% & 40.0\% & 0.00\% & 20.0\% & 0.0\%\tabularnewline
\bottomrule
\end{tabular}
\par\end{centering}
\caption{Percentage of all first ranks (wins) separated by dataset type: synthetic, motion sensors, sensor readings and image outlines. Rows may not add up to 100\% due to shared first ranks.\label{tab:wins_by_dataset}}

\end{table*}

In this experiment we found that WEASEL performs well independent of the domain. We studied the individual accuracy of each method on each of the 85 different datasets, and also grouped datasets by domain to see if different methods have domain-dependent strengths or weaknesses. We used the predefined grouping of the benchmark data into four types: synthetic, motion sensors, sensor readings and image outlines. Image outlines result from drawing a line around the shape of an object. Motion recordings can result from video captures or motion sensors. Sensor readings are real-world measurements like spectrograms, power consumption, light sensors, starlight-curves or ECG recordings. Synthetic datasets were created by scientists to have certain characteristics. For this experiment, we only consider the non-ensemble classifiers. Figure~\ref{fig:Classification-accuracy-for} shows the accuracies of WEASEL (black line) vs. the six core classifiers (orange area). The orange area shows a high variability depending on the datasets. 

Overall, the performance of WEASEL is very competitive for almost all datasets.  The black line is mostly very close to the upper outline of the orange area, indicating that WEASEL's performance is close to that of its best competitor. In total WEASEL has $36$ out of $85$ wins against the group of six core classifiers. On $69$ ($78$) datasets it is not more than $5\%$ ($10\%$) to the best classifier. The no-free-lunch-theorem implies that there is no single classifier that can be best for all kinds of datasets. Table~\ref{tab:wins_by_dataset} shows the correlation between the classifiers and each of the four dataset types. It gives an idea of when to use which kind of classifier based on dataset types. E.g., when dealing with sensor readings, WEASEL is likely to be the best, with $48.6\%$ wins. Overall, WEASEL has the highest percentage of wins in the  groups of sensor readings, synthetic and image outline datasets. Within the group of motion sensors, it performs equally good as LS and ST.

The main advantage of WEASEL is that it adapts to variable-length characteristic substructures by calculating discriminative features in combination with noise filtering. Thus, all datasets that are composed of characteristic substructures benefit from the use of WEASEL. This applies to most sensor readings like all EEG or ECG signals (CinC\_ECG\_torso, ECG200, ECG5000, ECGFiveDays, NonInvasiveFatalECG\_Thorax1, NonInvasiveFatalECG\_Thorax2, TwoLeadECG, ...), but also mass spectrometry (Strawberry, OliveOil, Coffee, Wine, ...), or recordings of insect wing-beats (InsectWingbeatSound). These are typically noisy and have variable-length, characteristic substructures that can appear at arbitrary time stamps~\cite{keoghtime}. ST also fits to this kind of data but, in contrast to WEASEL, is sensitive to noise. 

Image outlines represent contours of objects. For example, arrow-heads, leafs or planes are characterized by small differences in the contour of the objects. WEASEL identifies these small differences by the use of feature weighting. In contrast to BOSS it also adapts to variable length windows. TSBF does not adapt to the position of a window in the time series. ST and WEASEL adapt to variable length windows at variable positions but WEASEL also offers noise reduction, thereby smoothing the contour of an object.


Overall, if you are dealing with noisy data that is characterized by windows of variable lengths and at variable positions, which may contain superfluous data, WEASEL might be the best technique to use.

\subsection{Influence of Design Decisions on WEASEL's Accuracy}\label{influence}

\begin{figure}[t]
\includegraphics[width=1\columnwidth]{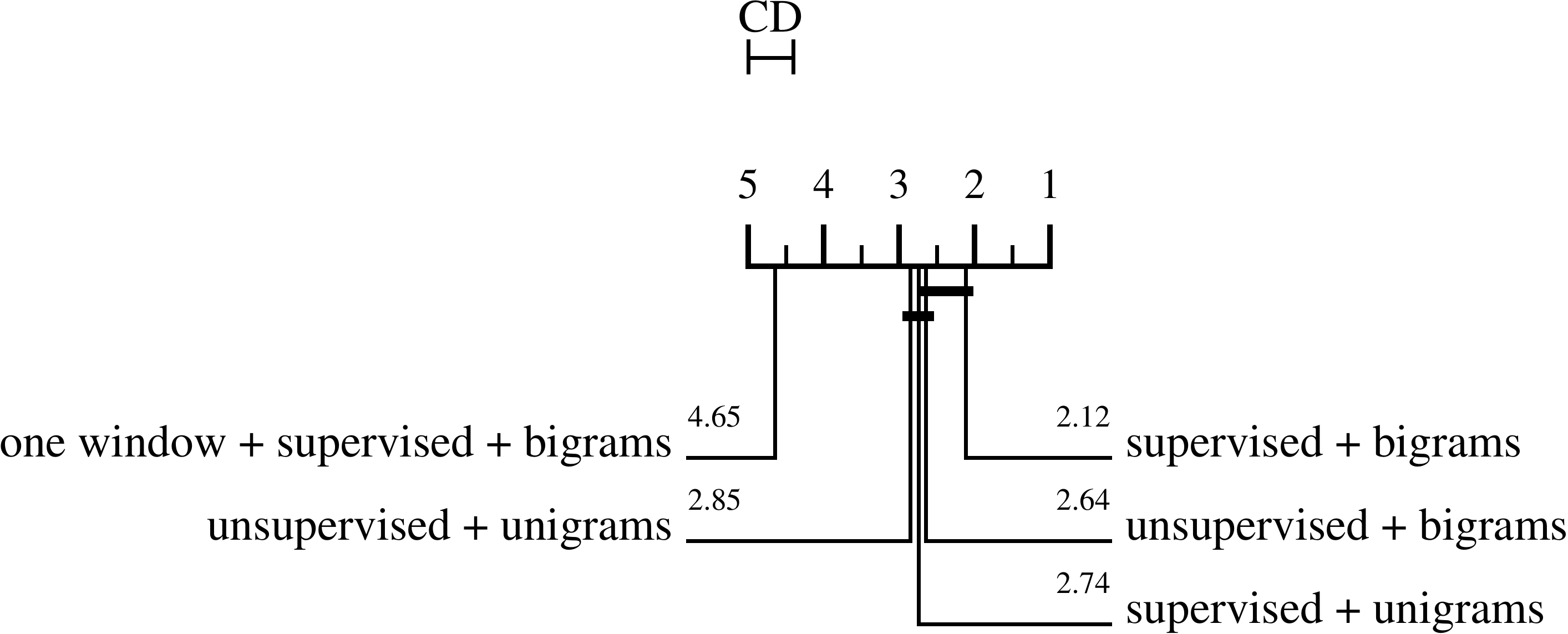}

\caption{Impact of design decisions on ranks. The WEASEL (supervised+bigrams) classifier has the lowest rank over all datasets.\label{fig:Impact-of-design}}

\end{figure}

We look into the impact of three design decisions on the WEASEL classifier: 
\begin{itemize}
\item The use of a novel supervised symbolic representation that generates discriminative features.
\item The novel use of bigrams that adds order-variance to the bag-of-patterns approach. 
\item The use of multiple window lengths to support variable length substructures.
\end{itemize}

We cannot test the impact of the Chi-Squared-test, as the feature space of WEASEL is not computationally feasible without feature selection (see Section~\ref{subsec:Chi-Squared-Test}).

Figure~\ref{fig:Impact-of-design} shows the average ranks of the WEASEL classifier where each extension is disabled or enabled: (a) "one window length, supervised and bigrams", (b) "unsupervised and unigrams", (c) "unsupervised and bigrams", (d) "supervised and unigrams", and (e) "supervised and bigrams". The single window approach is least accurate. This underlines that the choice of window lengths is crucial for accuracy. The unsupervised approach with unigrams is equal to the standard bag-of-patterns model. Using a supervised symbolic representation or bigrams slightly improves the ranks. Both extensions combined, significantly improve the ranks. 

The plot justifies the design decisions made as part of WEASEL. Each extension of the standard bag-of-patterns model contributes to the classifier's accuracy. Bigrams add order variance and the supervised symbolic representation produces disjoint feature sets for different classes. Datasets contain characteristic substructures of different lengths which is addressed by building a bag-of-patterns using all possible window lengths.

\subsection{Use Case: Smart Plugs}\label{use_case}
\begin{figure*}[t]
\subfloat[Accuracy vs prediction time for PLAID dataset.\label{fig:acsf1_predict}]{\includegraphics[width=1\columnwidth]{walltime_predict_plaid}}
\subfloat[Accuracy vs prediction time for ACS-F1 dataset.\label{fig:plaid_predict}]{\includegraphics[width=1\columnwidth]{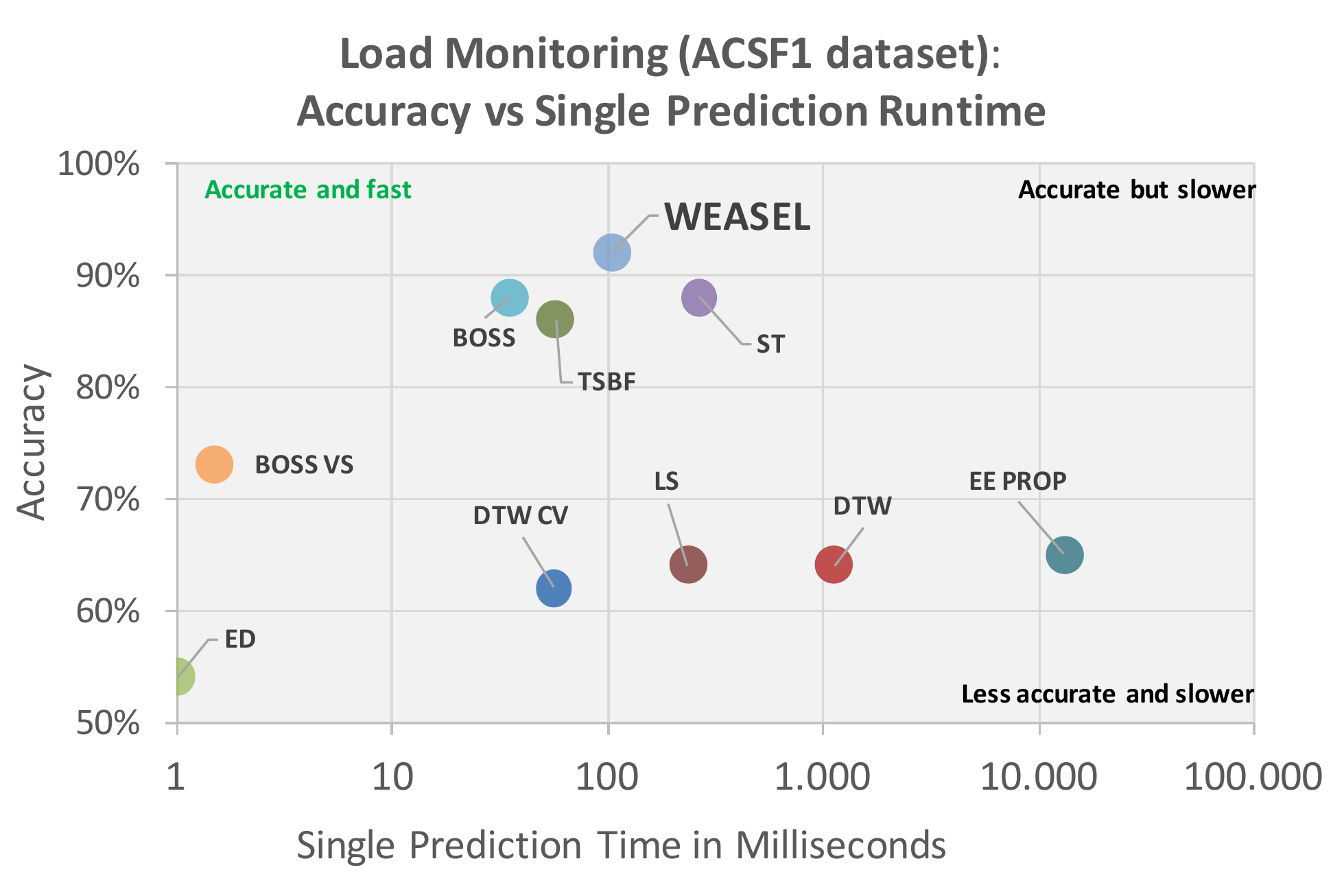}}

\subfloat[Accuracy vs train time for PLAID dataset.\label{fig:acsf1_train}]{\includegraphics[width=1\columnwidth]{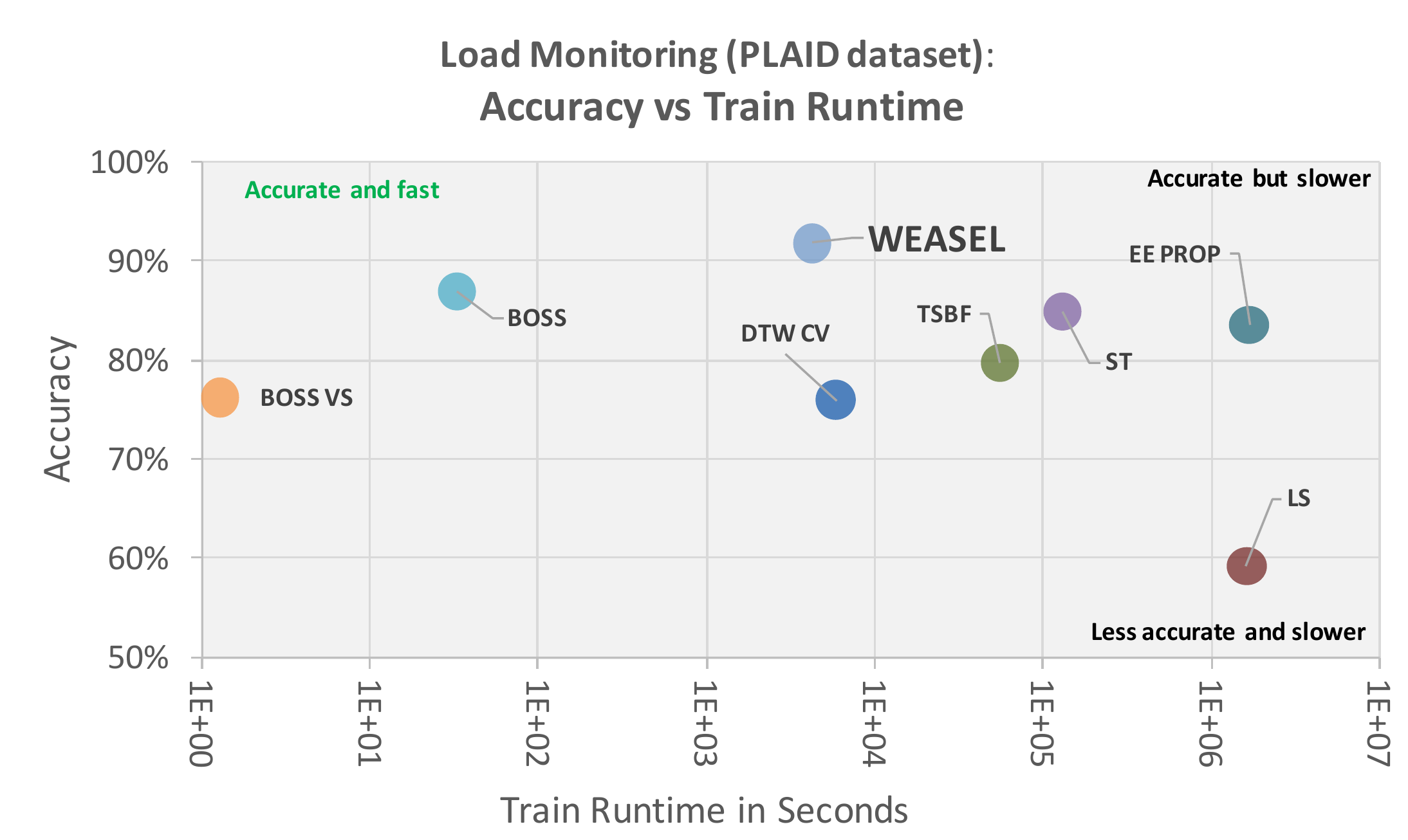}}
\subfloat[Accuracy vs train time for ACS-F1 dataset.\label{fig:plaid_train}]{\includegraphics[width=1\columnwidth]{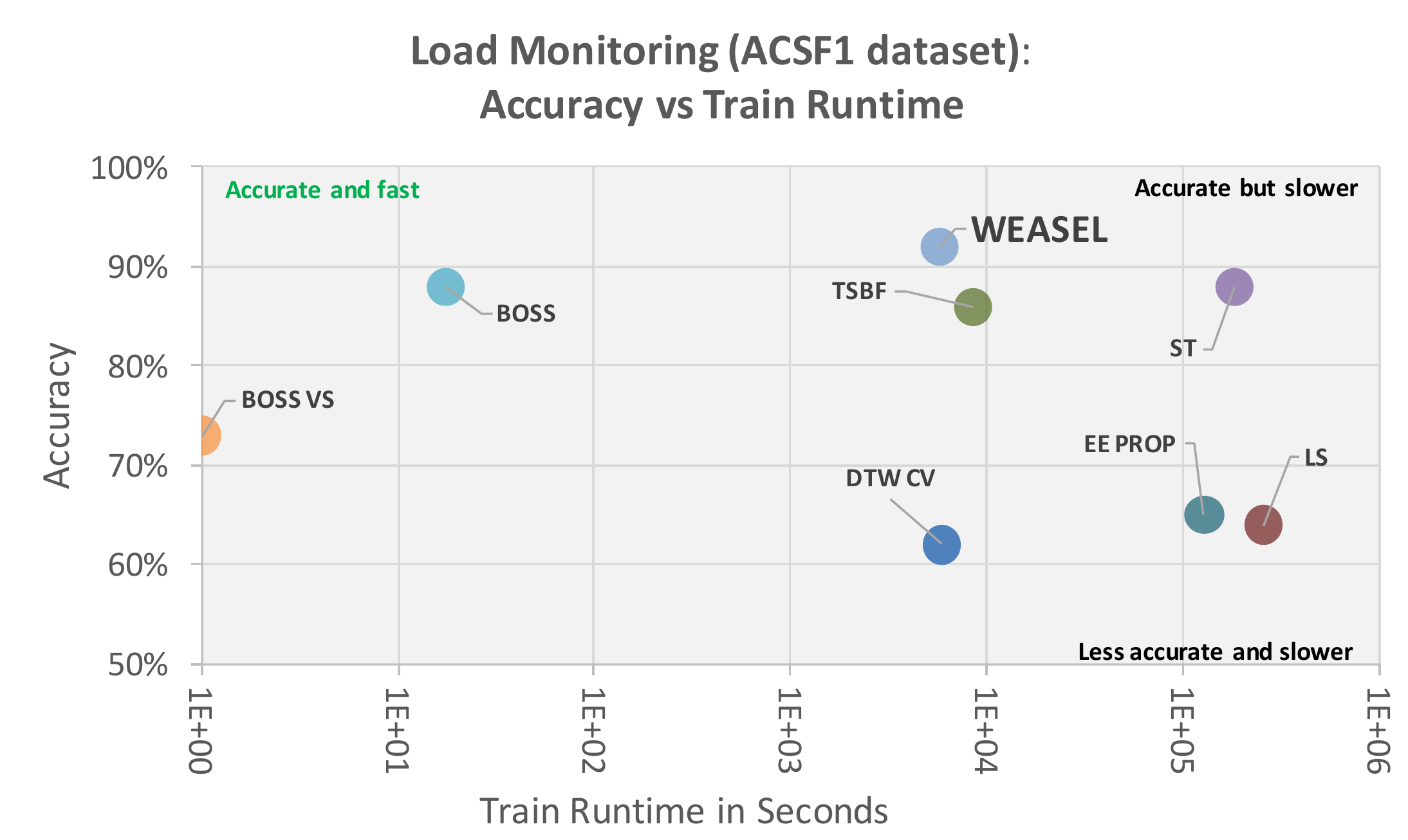}}

\caption{Use-Cases in Load Monitoring: PLAID and ACS-F1.\label{fig:ilm_datasets}}
\end{figure*}

Appliance load monitoring has become an important tool for energy savings~\cite{gisler2013appliance,gao2014plaid}. We tested the performance of different TSC methods on data obtained from intrusive load monitoring (ILM), where energy consumption is separately recorded at every electric device. We used two publicly available datasets ACS-F1~\cite{gisler2013appliance} and PLAID~\cite{gao2014plaid}. The PLAID dataset consists of 1074 signatures from 11 appliances. The ACS-F1 dataset contains 200 signatures from 100 appliances and we used their intersession split. These capture the power consumption of typical appliances including air conditioners, lamps, fridges, hair-dryers, laptops, microwaves, washing machines, bulbs, vacuums, fans, and heaters. Each appliance has a characteristic shape. Some appliances show repetitive substructures while others are distorted by noise. As the recordings capture one day, these are characterized by long idle periods and some high bursts of energy consumption when the appliance is active. When active, appliances show different operational states. 

Figure~\ref{fig:ilm_datasets} shows the accuracy and runtime of WEASEL compared to state of the art. COTE did not finish training after eight CPU weeks, thus we cannot report their results, yet. ED and DTW do not require training.

WEASEL scores the highest accuracies with $92\%$ and $91.8\%$ for  both datasets. With a prediction time of $10$ and $100$ ms it is also  fast. Train times of WEASEL are comparable to that of DTW CV and much lower than that of the other high accuracy classifiers.
 
On the large PLAID dataset WEASEL has a significantly lower prediction time than its competitors, while on the small sized ACS-F1 dataset the prediction time is slightly higher than that of DTW or BOSS. 1-NN classifiers such as BOSS and DTW scale with the size of the train dataset. Thus, for larger train datasets, they become slower. At the same time, for small datasets like PLAID, they can be quite fast.

The results show that our approach naturally adapts to appliance load monitoring. These data show how WEASEL automatically adapts to idle and active periods and short, repetitive characteristic substructures, which were also important in the sensor readings or image outline domains (Section~5.4).

Note that the authors of the ACS-F1 dataset scored $93\%$~\cite{ridi2013automatic} using a hidden Markov model and a manual feature set. Unfortunately their code is not available and the runtime was not reported. Our accuracy is close to theirs, while our approach was not specially adapted for the domain. 

\section{Conclusion and Future Direction}

In this work, we have presented WEASEL, a novel TSC method following the bag-of-pattern approach which achieves highly competitive classification accuracies and is very fast, making it applicable in domains with very high runtime and quality constraints. The novelty of WEASEL is its carefully engineered feature space using statistical feature selection, word co-occurrences, and a supervised symbolic representation for generating discriminative words. Thereby, WEASEL assigns high weights to characteristic, variable-length substructures of a TS. In our evaluation on altogether 87 datasets, WEASEL is consistently among the best and fastest methods, and competitors are either at the same level of quality but much slower or equally fast but much worse in accuracy. 

In future work, we will explore two directions. First, WEASEL currently only deals with univariate TS, as opposed to multi-variate TS recorded from an array of sensors. We are currently experimenting with extensions to WEASEL to also deal with such data; a first approach which simply concatenates the different dimensions into one vector shows promising results, but requires further validation. Second, throughout this work, we assumed fixed sampling rates, which let us omit time stamps from the TS. In future work, we also want to extend WEASEL to adequately deal with TS which have varying sampling rates.

\bibliographystyle{abbrv}
\bibliography{boss}

\end{document}